%% file: main_v4.1_jor.tex
\documentclass[%
 aip,
 amsmath,amssymb,
 reprint,%
]{revtex4-2}

\usepackage{graphicx}% Include figure files
\usepackage{dcolumn}% Align table columns on decimal point
\usepackage{bm}% bold math
\usepackage[utf8]{inputenc}
\usepackage[T1]{fontenc}
\usepackage{mathptmx}
\usepackage{etoolbox}

\usepackage{textcomp}
\usepackage{natbib,setspace}
\bibpunct[, ]{[}{]}{;}{n}{,}{,}
\usepackage{jabbrv}
\usepackage{enumitem}
\usepackage{layouts}
\usepackage{soul}
\usepackage{float}
\usepackage{placeins}
\usepackage{appendix}
\usepackage{xcolor}
\definecolor{sysAnna}{HTML}{0077BB}
\definecolor{sysClasen}{HTML}{EE7733}
\definecolor{sysCalabrese}{HTML}{AA2277}
\definecolor{sysGaillard}{HTML}{009988}
\definecolor{sysHuPSDOP}{HTML}{5D3A9B}
\definecolor{sysHuPEO}{HTML}{C1121F}
\definecolor{sysHuPIB}{HTML}{8C5C0A}
\newcommand{\legdot}[1]{\textcolor{#1}{\ensuremath{\bullet}}}
\usepackage{nicefrac,xfrac}
\usepackage[Euler]{upgreek}
\usepackage{comment}

\input{./macros_velbgeometry.tex}

\makeatletter
\def\@email#1#2{%
 \endgroup
 \patchcmd{\titleblock@produce}
 {\frontmatter@RRAPformat}
 {\frontmatter@RRAPformat{\produce@RRAP{*#1\href{mailto:#2}{#2}}}\frontmatter@RRAPformat}
 {}{}
}%
\makeatother

\begin{document}

\title{To win, a model must thin: Capillary thinning as a benchmark complex flow for constitutive models of viscoelastic polymer solutions}

\author{Ranganathan Prabhakar}
\affiliation{Department of Mechanical and Aerospace Engineering, Monash University, Clayton, Victoria, Australia}
\email{prabhakar.ranganathan@monash.edu}

\author{Joseph P. Connell}
\affiliation{Department of Mechanical and Aerospace Engineering, Monash University, Clayton, Victoria, Australia}

\date{\today}

\begin{abstract}

Capillary thinning of a liquid bridge is an exemplar of complex flow, where the macroscopic geometry couples tightly to the microscopic evolution of polymer conformations. Since its inception, capillary-breakup rheometry (CBR) has been viewed as a tool for measuring a single  relaxation time.  Yet experiments show that the apparent relaxation time depends systematically on polymer concentration, device geometry, and the preparation protocol. We argue that this variability is not a flaw, but evidence that thinning should be treated as a benchmark complex flow for testing constitutive models. We recast the output of a CBR experiment as the self-selected elastic strain rate, expressed through the elastic Weissenberg number \(\Wie\), rather than an apparent relaxation time, and organize it in an elastocapillary Pipkin diagram --- \(\Wie\) against a geometry-controlled Deborah number. A single-mode, mid-filament stress balance yields a family of Pipkin curves with universal features --- a low-\(\Deo\) plateau and a finite-extensibility-constrained rise --- that a scaling analysis collapses onto a master curve, with an elastic-onset-referenced Deborah number absorbing the unmeasured initial prestretch. The Conformation- and Concentration-Dependent Drag (C2D2) model, acting through coil--stretch hysteresis, lowers the plateau below the Entov--Hinch value and organizes data spanning decades in molecular weight and concentration and a range of devices, where the classical FENE-P model cannot. The Pipkin diagram framework offers a path toward master plots for classes of polymer solutions, clarifying what is universal in extension-dominated flows. 
\end{abstract}

\maketitle

\section{\label{sec:intro} Introduction}

In complex flows of complex fluids, the macroscopic evolution of the flow field is strongly coupled to the microscopic evolution of the underlying microstructure. To eventually be useful in engineering design, microstructure-based constitutive models of polymer solutions must be judged by their ability to predict this micro--macro feedback and, through it, the experimentally observed behaviour of viscoelastic fluids in contraction flows, cylinder wakes, jets, and other strongly inhomogeneous settings. In many such flows, the most important regions are those experiencing strong extensional straining, since polymer stretching and the resulting normal stresses are often the dominant mechanisms through which viscoelasticity manifests itself.

Capillary thinning of a liquid bridge belongs to this class of micro--macro-coupled flows. The bridge geometry dictates the evolving interaction between surface tension, flow, and polymeric stresses, and thereby sets up a predominantly extensional flow at the necking plane. At the same time, capillary thinning occupies a special place in rheology because it lies at the boundary between a complex flow and a rheometric flow.

The gold standard in extensional rheometry is the filament-stretching extensional rheometer (FiSER), developed by Sridhar and co-workers and others \citep{Tirtaatmadja1993, gupta, mckinleysridhar, Tuladhar2008-nq}. In FiSER, capillary thinning is exploited together with actively controlled end-plate motion to maintain a prescribed, nearly homogeneous extension rate at the neck of a viscoelastic sample. The end-plate forces are measured using force transducers, allowing the normal stress difference to be extracted at the same location where the strain rate is known. In this sense, FiSER realizes the conditions required of a rheometric flow and has become a key tool for testing and validating constitutive models for dilute and nondilute polymer solutions as well as polymer melts. Its operation, however, is experimentally demanding, typically requiring sophisticated feedback control, force measurement, and sufficiently viscous samples.

In recent years, a simpler class of experiments has become increasingly popular. Instead of continually adjusting the end-plate motion, as in FiSER, to maintain a prescribed strain rate at the neck, one simply creates a sufficiently long liquid bridge and allows it to thin spontaneously toward breakup. The subsequent neck-radius decay is then measured, and extensional rheological properties are inferred by fitting constitutive or asymptotic model equations to the observed radius--time data \citep{bazilevsky1, liang, McKinley2000, EvalInkJet}. This broad approach, which we refer to here as capillary-breakup rheometry (CBR), avoids the need for expensive force transducers and sophisticated feedback control, and can often be implemented as a comparatively simple benchtop technique. Besides Capillary Breakup Extensional Rheometers (CaBER) devices that employ mechanically separated end-plates, other variants such as the Acoustically Driven Microfluidic Extensional Rheometer (ADMiER) \citep{Bhattacharjee2011, mcdonnell2015} and the Dripping-on-Substrate (DoS) device \citep{Dinic2015-fh, Robertson_Calabrese, Zhang_Calabrese} generate the initial liquid bridge by different means.

Unlike a conventional rheometer, however, these CBR devices do not seek to impose a strain rate at the neck. Instead, once the bridge has become sufficiently slender, a nearly cylindrical filament develops locally in which the extensional strain rate is approximately homogeneous. The thinning neck imposes an unsteady capillary stress on the fluid, and the self-selected strain rate at the neck emerges as the measurable output. Remarkably, early experiments showed that in polymer solutions the self-selected strain rate in the elastic stage of thinning is often nearly constant \citep{bazilevsky1, liang}.

Interest in this method was ignited by the seminal analysis of \citet{entovhinch}, who coupled a multimode constitutive model for a viscoelastic polymer solution to the macroscopic stress balance at the necking plane. Their analysis revealed two key findings. First, when the polymeric contribution dominates over the solvent-viscous contribution, the neck enters an elastic regime in which the radius decays nearly exponentially with time, corresponding to a nearly constant Hencky strain rate. Second, this self-selected elastic strain rate, \(\edote\), was predicted to be simply related to the largest relaxation time \(\lamI\) of the discrete relaxation spectrum used in the model: the Weissenberg number corresponding to \(\edote\),
\begin{gather}
\Wie = \lamI \edote = \frac{2}{3}.
\end{gather}

This understandably created considerable excitement in the rheological community. It suggested that one may convert the experimentally-observed strain-rate in the elastic regime into a relaxation time,
\begin{gather}
\lame = \frac{2}{3\edote},
\end{gather}
and that, if the Entov-Hinch (EH) prediction were true,  $\lame = \lamI$, the intrinsic relaxation time of the polymer solution. This meant that a comparatively simple and inexpensive experiment could be used to estimate the largest relaxation time of a polymer solution directly from the slope of the exponential neck-radius decay on a semilogarithmic plot, thereby circumventing more elaborate rheometric protocols based on shear measurements. The extracted time scale could then be used for further characterization, for example in estimating the molecular weight of the polymer solute. Following the EH study, the first set of experiments by \citet{annamckinley} on polystyrene (PS) Boger fluids, for which $\lamI$ were estimated independently, appeared to support the prediction that $\lame = \lamI$.

Subsequent experiments, however, have shown that \(\lame \neq \lamI\) in general. \citet{clasenetal} demonstrated this across PS solutions of various molecular weights, concentrations, and solvents, measured in two different CaBER devices, against independent values of \(\lamI\) obtained from small-amplitude oscillatory shear (SAOS) experiments. The measured \(\lame\) further displays a strong and systematic concentration dependence, quite different from the much weaker concentration dependence of \(\lamI\) in dilute solutions. More recently, \citet{Gaillard2024-xu} cast further doubt on whether \(\lame\) extracted from CBR thinning experiments can be interpreted as an intrinsic material property at all. Using the same polymer solution in devices with different plate radii, they obtained systematically different values of the apparent relaxation time. Since these values vary reproducibly with the geometry of the device, \(\lame\) cannot be regarded as a material property. A further question concerns the effect of the initial deformation of the polymer during the creation of the liquid bridge. As noted earlier, different CBR devices use different methods to create and stabilize the liquid bridge before capillary thinning commences. To probe the effect of the bridge-creation step, \citet{Aisling2024-ie} formed liquid bridges using mechanically separated plates across a range of plate-separation rates. They showed that the active stretching phase prior to passive thinning influences the late-stage thinning rate, and that a sufficiently large initial extension rate is required before that thinning rate ceases to depend on the imposed protocol. \citet{Gaillard2025-on} likewise show that the neck radius at the onset of polymer-controlled thinning depends on whether the chains have been actively prestretched or are entering the bridge close to equilibrium. \citet{Hu2025-rt} trace the systematic dependence of \(\lame\) on device scale to prestretching generated within the thinning process itself: integrating the FENE-P model forward from equilibrium, they show that chains stretched during the early viscocapillary stage enter the elastic stage at a stretch that grows with the initial filament diameter, so that the apparent thinning rate is faster than that implied by the relaxation time of the underlying model. Their measurements extend the evidence for this scale dependence across dripping, dripping-onto-substrate, and CaBER devices and three polymer--solvent chemistries; we fold their datasets into the comparisons presented here, and defer a detailed discussion of their proposed correction of \(\lame\) --- and of how our reading of the same phenomenon differs --- to Sec.~\ref{sec:discussion}. Taken together, these studies suggest that the strain history experienced by the chains before the elastic stage commences --- whether imposed by the bridge-creation protocol or accumulated during the early thinning itself --- also influences \(\lame\).

In the title of their paper, \citet{Gaillard2024-xu} counsel that one should ``beware of CaBER''. We take a different view: the evidence does not invalidate the experiment but the theory the experiment was designed to test. Recalling the original analysis of \citet{entovhinch}, one observation is in fact confirmed by the data: at polymer concentrations sufficient to produce a discernible elastic regime (though not necessarily so large that the solution leaves the dilute regime) the neck radius
decays exponentially in time at a nearly steady strain rate \(\edote\). What fails is the second prediction, namely, \(\Wie = 2/3\). The problem therefore lies not with CBR but with the constitutive assumptions on which that prediction rests; CBR remains valid as a probe for testing constitutive models.

In general, this use of rheometry to test constitutive models has a long and distinguished history. In recent times, FiSER experiments have been indispensable in testing microstructure-based models of entangled polymer solutions and melts. What we wish to argue here is that CBR provides something more than a standard rheometric test. Because the strain rate at the neck is self-selected through a two-way coupling between the evolving polymer microstructure and the macroscopic flow of the bridge, CBR is closer in character to canonical benchmark complex flows -- the 4:1 contraction, flow past a cylinder, serpentine channel flow -- than to a homogeneous rheometric flow where either the strain-rate or stress is carefully controlled. In benchmark flows and in CBR, one sets the initial and boundary conditions and allows the flow to evolve. CBR is, however, special among these: the dynamics localize to the necking plane, so the governing equations are zero-dimensional in space and the micro--macro coupling is preserved without the additional complexity of spatial inhomogeneity. CBR can thus be viewed as perhaps the ``simplest complex flow'' one can impose on a polymer solution -- and, as such, a vital gateway. A constitutive model that cannot describe CBR is unlikely to do better in the spatially inhomogeneous flows for which it is ultimately needed. To win, it must first thin correctly.

With this perspective, a simple shift in interpretation is natural: we treat the ``output'' of a CBR device as the self-selected strain rate \(\edote\), rather than the apparent relaxation time \(\lame\). This shift immediately removes the expectation that the CBR output is an intrinsic material property, and recasts the interpretation problem as one of \emph{predicting} \(\edote\) from the molecular parameters of the polymer solution and the initial and boundary conditions of the flow. We aim to show that the liquid-bridge geometry and preparation-protocol dependences of \(\edote\) can be explained in just this way, and that CBR measurements across four experimental systems -- those of \citet{annamckinley} and \citet{clasenetal} on PS Boger fluids, \citet{Calabrese2025-re} on PS-in-dioctylphthalate (DOP) solutions, and \citet{Gaillard2024-xu} on polyethyleneoxide (PEO) solutions in a viscous PEG--water solvent -- augmented by the recent multi-device datasets of \citet{Hu2025-rt} on PS-DOP, PEO--PEG, and PIB--polybutene solutions, prove largely mutually consistent. Doing so requires a better microstructure-based constitutive model than conventional descriptions for dilute polymer solutions such as the Oldroyd-B (Hookean dumbbell) or finitely extensible nonlinear elastic--Peterlin (FENE-P) models.

The missing ingredient in these conventional models is the conformation dependence of intra- and intermolecular hydrodynamic interactions (HI). In the FENE-P model, the polymer chain is represented as a single dumbbell whose elastic response combines backbone flexibility and entropic resistance with a finite extensibility imposed at large stretch. This is a deliberate, physically defensible reduction, designed to capture the dominant resistance to chain extension while sidestepping the full chain dynamics. The model handles HI in two ways, both kept fixed throughout the flow. Intramolecular HI is absorbed into the choice of relaxation time: the FENE-P time scale is taken to be the longest Zimm relaxation time of an equilibrium coil (in multimode descriptions, the full Zimm relaxation spectrum is used). Intermolecular HI is assumed negligible in dilute solutions, \(c/\cstar < 1\), where the chains are typically far apart. However, to explain the strong concentration dependence of \(\lame\) observed in their experiments with dilute solutions, \citet{clasenetal} pointed out that conventional dilute-solution models neglect the fact that stretched polymer chains hydrodynamically interact more strongly with one another than quiescent coils do. A solution that behaves as dilute under equilibrium conditions, with weak intermolecular interactions and only weak concentration dependence of its rheological properties, may therefore behave quite differently under strong extensional flow. In that sense, dilute polymer solutions may effectively ``self-concentrate'' as chains stretch.

This idea led to the development of the Conformation- and Concentration-Dependent Drag (C2D2) model by Prabhakar and co-workers \citep{Prabhakar2016, Prabhakar2017-ri}. The C2D2 model extends the conventional FENE-P description by using blob-based arguments \citep{rubinsteincolby} to account for the effect of intra- and intermolecular HI on the average friction coefficient, and hence on the largest relaxation time, of partially stretched polymer chains. The model successfully predicted the concentration dependence of \(\Wie\) observed by \citet{clasenetal}. That work further showed that the selected elastic Weissenberg number is closely related to coil-stretch hysteresis, and in particular to the critical stretch-to-coil transition value \(\Wisc\), which is smaller than the coil-to-stretch transition value \(\Wics=1/2\). Indeed, comparison with CBR data led to the intriguing conclusion that the ratio \(\Wics/\Wisc\), which measures the width of the hysteresis window, increases with concentration in dilute solutions, peaks near the overlap concentration \(\cstar\), and then decreases again as the solution enters the semidilute regime. This prediction of the C2D2 model was qualitatively verified using FiSER experiments and Brownian Dynamics simulations of unentangled polymer solutions \citep{Prabhakar2017-ri}. 

We show here that the C2D2 model is central to reconciling the diverse CBR experimental observations. We further show that the seemingly disparate liquid-bridge geometry dependences of \(\edote\) reported across these studies -- most strikingly the device-geometry sweep of \citet{Gaillard2024-xu} -- collapse onto a single master plot. We proceed in stages. The modeling framework -- a single-mode mid-filament stress balance for a slender liquid bridge, coupled to either the FENE-P or the C2D2 description of polymer conformation evolution -- is set out in Sec.~\ref{sec:model}. Using the conventional FENE-P model, we first show that the geometric parameters of the liquid bridge influence the stress balance through a Deborah number, and that \(\edote\) -- equivalently \(\Wie\) -- depends systematically on the bridge geometry through it. To organize these predictions we introduce the \textit{elastocapillary Pipkin diagram}, in which the \(\Wie\) output of a CBR experiment is plotted against the Deborah-number input, yielding a family of \textit{Pipkin curves}, one for each polymer concentration. A scaling analysis -- which revisits the Entov--Hinch argument from the perspective of transient trajectories in the extensional-viscosity--Weissenberg-number phase space -- organizes these curves and collapses them onto a master curve, and the conformation- and concentration-dependent drag of the C2D2 model, acting through coil-stretch hysteresis, proves essential to reconciling the experimental data; along the way we propose a redefinition of the Deborah number that absorbs an unmeasured initial prestretch into a single observable. The collapse is not achieved by the predicted factors alone, however: each system needs a single empirical shift of its corner, and the magnitude of that shift is itself diagnostic of physics the model does not yet capture. These predictions, together with their comparison against experimental CBR data across the seven datasets introduced above, are presented in Sec.~\ref{sec:results}. Section~\ref{sec:discussion} then takes up the implications of the framework -- the interpretation of CBR data, the role of capillary thinning as a benchmark complex flow for testing constitutive models, and the inverse problem of inferring molecular parameters from CBR data -- and Sec.~\ref{sec:concl} summarizes our conclusions.

% [Intro roadmap merged into the overview paragraph above, 2026-06-13; subsection-level breakdown now lives in the Results & Discussion opener.]

\section{\label{sec:model} Modeling}

We consider polymer solutions in a Newtonian solvent of viscosity \(\etas\), density \(\rho\), and surface tension coefficient \(\gamma\) (with air), at absolute temperature \(T\). The polymer concentration is given by the number density \(c\), and the chains are assumed monodisperse with contour length \(L\) and equilibrium coil size estimated by the root-mean-squared end-to-end distance \(\Reo\). The critical overlap concentration \(\cstar\) is that at which the screening of intramolecular HI by neighbouring chains becomes significant. We estimate it here as
\begin{gather}
\cstar = \Reo^{-3},
\end{gather}
up to an \(O(1)\), possibly non-universal, prefactor that could in principle be determined for a given system as the ratio of its measured overlap concentration to \(\Reo^{-3}\). Here, however, we set this prefactor to unity and define the reduced concentration
\begin{gather}
\phi = \frac{c}{\cstar}.
\end{gather}
When comparing with experiment, for the two datasets that probe concentration effects we estimate \(\cstar\) independently from the measurements, form the corresponding experimental \(\phi\), identify it with the model's reduced concentration, and run the model with the prefactor held at unity; the calibration of the prefactor is deferred to future refinements of the model. We focus primarily on nominally dilute solutions, \(\phi \le 1\), although the model also yields predictions for \(\phi>1\). The longest relaxation time under quiescent conditions at a given concentration is denoted \(\lamI\).

The microscopic description of the chains is simplified to retain only the features most relevant to dilute and weakly semidilute solutions near the \(\theta\) condition. Each polymer is modeled as a flexible random walk of
\begin{gather}
\NK = \frac{L^2}{\Reo^2} \gg 1
\end{gather}
Kuhn segments, where \(L\) is the contour length and \(\Reo\) the equilibrium coil size. Excluded-volume and electrostatic interactions are neglected, so that intra- and intermolecular couplings are limited to backbone connectivity, chain flexibility, and HI. The strength of intramolecular HI is characterized by the dimensionless Kuhn-segmental parameter
\begin{gather}
\hsK = \frac{a_\textsc{k}}{\sqrt{\pi}\,b_\textsc{k}},
\end{gather}
where \(b_\textsc{k}\) is the Kuhn length and \(a_\textsc{k}\) is the hydrodynamic radius of the Kuhn segment. This setting defines the scope of the models developed below: non-polyelectrolyte polymer solutions in solvents close to \(\theta\) conditions, where backbone elasticity and HI dominate. Thus, the dimensionless parameters that characterize the polymer solution are $\phi$, $\NK$ and $\hsK$.

In a typical capillary-thinning experiment, a liquid bridge of volume \(\Vb\) is created between two circular end-plates of radius \(\Rb\) separated by a distance \(\Lb\). Once formed, the bridge thins under surface tension, and the minimum neck radius \(R(t)\) decays with time \(t\). The local strain rate at the necking plane is then
\begin{gather}
\dot{\varepsilon}(t) = -\,2\,\frac{d\ln R}{dt},
\end{gather}
corresponding to the uniaxial extensional flow established at the neck.

The initial stage of thinning involves transients that depend on the overall bridge geometry and the method of formation. In Newtonian liquid bridges, as thinning proceeds the neck becomes sufficiently slender that the global dimensions of the bridge no longer influence the local dynamics. The evolution then approaches a well-studied self-similar regime in which the neck radius depends only on material parameters and surface tension forces \citep{eggers1997, papageorgiou1995, li_sprittles_2016}. We therefore take the onset of self-similar thinning of the corresponding pure-solvent bridge as the initial time \(t=0\), with neck radius \(R_0\). For the moment, we assume that during this stage the polymer chains remain close to equilibrium, so that polymer stresses are negligible. We shall later consider the effect of chains being partially stretched at $t = 0$. 

We assume that gravitational effects are negligible \emph{i.e.} that the Bond number, $\Bo =  \rho g R_0^2/\gamma$, is small. The initial solvent-dominated dynamics are characterized by the Ohnesorge number,
\begin{gather}
\Oh = \frac{\etas}{\sqrt{\rho \gamma R_0}},
\end{gather}
which compares viscous to inertio-capillary effects. The viscous-solvent regime of interest here is attained when \(\Oh \gg 1\), though in practice \(\Oh \gtrsim 1\) suffices \citep{li_sprittles_2016}; we restrict attention to this regime throughout. The initial self-similar evolution of the neck for \(t>0\) is then characterized by the visco-capillary time scale
\begin{gather}
\tcap = \frac{\etas R_0}{\gamma},
\label{e:timescales}
\end{gather}
during which the neck radius decays linearly in time \citep{li_sprittles_2016}.

In dilute polymer solutions, the initial decay of the neck radius closely follows that of the pure solvent while polymer stresses are still weak. After a period of order the capillary timescale \(\tcap\), the chains are sufficiently stretched that their stresses become comparable to capillarity, marking the onset of qualitatively different behaviour from the pure solvent. We refer to the regime in which the polymer contribution to the stress dominates over the solvent-viscous contribution as the \textit{elastic} regime, and characterize the dynamics there through the unsteady strain rate at the neck, written in dimensionless form as the Weissenberg number
\begin{gather}
\Wi(t) = \lamI \,\dot{\varepsilon}(t).
\label{e:widef}
\end{gather}

This is crucial because, as we shall show in detail, the coil-stretch transition and the associated hysteresis in the \emph{steady-state} extensional viscosity play an important role in determining the dynamics in the elastic regime. As is well known, the coil-to-stretch transition for initially coiled chains is triggered only after \(\Wi\) exceeds the critical value \(\Wics = 1/2\). Moreover, in order for stretched chains to maintain their extension and sustain large polymeric stresses, \(\Wi\) must also remain above the critical \emph{stretch-to-coil transition} value, \(\Wisc\). The conventional FENE-P model does not predict a steady-state coil-stretch hysteresis \emph{i.e.} it predicts $\Wisc = \Wics$. With conformation-dependent drag, $\Wisc < \Wics$, and this difference between the two models is central to understanding why the FENE-P model does not predict the concentration dependence of \(\Wie\) observed in CBR experiments.

Our principal aim is to understand how the Weissenberg number selected in this elastic regime depends on the geometry of the liquid bridge. Geometry enters through its influence on the initial neck radius \(R_0\): the plate radius \(\Rb\), plate spacing \(\Lb\), and sample volume \(\Vb\) together fix \(R_0\), which sets the capillary timescale and hence the Deborah number, the ratio of the longest relaxation time \(\lamI\) to that timescale,
\begin{gather}
\Deo = \frac{\lamI}{\tcap} = \frac{\lamI\gamma}{\etas R_0},
\label{e:dedef}
\end{gather}
built on the initial neck radius \(R_0\).

\subsection{\label{sec:mfsbe} Stress balance at the neck}

The macroscopic flow at the neck plane is governed by a balance of viscous, capillary, and polymeric stresses. Collecting the (solvent) viscous contribution on the left and the capillary and polymeric contributions on the right, the balance of first normal stresses at the necking plane is approximately \citep{entovhinch, McKinley2000,annamckinley,Tirtaatmadja2006}
\begin{gather}
-\,\frac{6\,\etas\,\dot R}{R}
\;=\;
\frac{(2X-1)\,\gamma}{R}\;-\;\NIp,
\label{e:fullbal}
\end{gather}
where \(R(t)\) is the mid-filament radius (\(\dot R<0\) while the neck thins), \(\etas\) is the solvent viscosity, \(\gamma\) the surface tension, and \(\NIp\) the (dimensional) polymer first normal-stress difference. Here \(X=F_\mathrm{b}/(2\pi\gamma R)\) is the ratio of the tensile force \(F_\mathrm{b}\) exerted at the boundary by the end plates to the local capillary force \citep{McKinley2000, Gaillard2025-on}. In general \(X\) evolves through the thinning; during the self-similar pre-elastic stage of the pure solvent it approaches a constant. This is discussed further below.

We scale the radius by the initial neck radius \(R_0\) and stresses by \(\gamma/R_0\), write the strain rate at the neck as the Weissenberg number \(\Wi=-2\,\lamI\dot R/R\), and normalize the polymer normal stress by the modulus, \(\tNIp=\NIp/\ckBT\) (given later in terms of the dimensionless conformation in Sec.~\ref{sec:conf}). Taking the viscocapillary time \(\tcap\) as the macroscopic timescale, the balance \eqref{e:fullbal} rearranges to give the Weissenberg number at the neck,
\begin{gather}
\Wi = \frac{1}{3}\left[\frac{(2X-1)\,\Deo}{R} - \UI\,\phi\,\tNIp\right],
\label{e:Wi-visc}
\end{gather}
with the polymer prefactor
\begin{gather}
\UI = \frac{\lamI \kBT}{\etas \Reo^{3}}.
\label{e:U1def}
\end{gather}
The prefactor \(\UI\) is the ratio of the relaxation time \(\lamI\) to the coil diffusion time \(\etas\Reo^{3}/\kBT\). It approaches the parameter-free value \(0.325\) as \(\NK\to\infty\) \citep{doiedw}, with a weak non-universal dependence on \(\hsK\) otherwise \citep{Prabhakar2016,Prabhakar2017-ri,SI}. Just as the geometric Deborah number sets the weight of the polymer term relative to capillarity, it is the device geometry, entering through \(\Deo\), that controls the strength with which the polymer stress couples to the macroscopic balance.

\paragraph*{The \(X\) factor.}
The dimensionless end-plate tension $X$ varies with time. During the initial self-similar pre-elastic thinning, it is approximately constant and sets the initial Weissenberg number at \(t=0\) to be proportional to the Deborah number,
\begin{gather}
\Wio = \Hv\,\Deo.
\label{e:Wio}
\end{gather}
In the self-similar viscocapillary thinning of interest here, \(X = \Xv=0.7127\) \citep{papageorgiou1995, McKinley2000, li_sprittles_2016}, so that \(\Hv=(2\Xv-1)/3=0.142\). Once polymer stresses dominate, the force factor crosses over from its pre-elastic value \(\Xv\) to the elastic value \(\Xe=3/2\) \citep{Eggers2020-rn, Deblais2020-of}, a switch the model effects smoothly \citep{SI}.

\subsection{\label{sec:conf} Polymer conformation and stress}

The conformation tensor \(\bM\) represents the average chain shape. Its trace gives the instantaneous stretch ratio
\begin{gather}
M = \tr(\bM) = \frac{\Ree^2}{\Reo^2},
\end{gather}
where \(\Ree\) is the instantaneous mean end-to-end distance. In the C2D2 model, \(\bM\) evolves according to
\begin{equation}
\frac{d \bM}{d t}
=
\bkappa \cdot \bM
+
\bM \cdot \bkappa^\textsc{t}
-
\frac{1}{\Deo \,\nu}
\left(
f\,\bM - \dfrac{1}{3}\ut
\right),
\label{e:confevol}
\end{equation}
where \(\bkappa\) is the dimensionless rate-of-strain tensor associated with the local extensional flow, \(\ut\) is the identity tensor, and time has been scaled by the macroscopic viscocapillary timescale \(\tcap\). The first two terms describe affine stretching and orientation by the flow, while the last term represents relaxation toward equilibrium.

The entropic restoring force in a stretched polymer chain, of stiffness \(\kBT/\Reo^2\), is treated using the FENE force law with the Peterlin closure \citep{Peterlin1966b, dpl2}. The dimensionless stiffening factor due to finite-extensibility is obtained in this model as
\begin{equation}
f = \frac{\NK - 1}{\NK - M},
\label{e:FEf}
\end{equation}
which diverges as \(M\) approaches \(\NK\), corresponding to chains approaching full stretch.

The function \(\nu(M; \phi, \NK, \hsK)\) in Eq.~\eqref{e:confevol}, which is central to the C2D2 model, represents the conformation- and concentration-dependent drag ratio, namely the ratio of the average drag coefficient \(\zeta\) of partially stretched chains to its equilibrium value \(\zetao\); thus \(\nu=1\) at equilibrium and increases as chains stretch or as concentration rises. Its dependence on \(M\), \(\NK\), \(\hsK\), and \(\phi\) is obtained from blob-based scaling arguments that account for the attenuation of intramolecular HI as chains are stretched, and for the screening of those interactions by neighbouring chains \citep{Prabhakar2016, Prabhakar2017-ri, c2d2report}. These effects are neglected in the conventional FENE-P model, which is recovered by setting \(\nu=1\) at all times.

With \(\bM\) determined at any instant by integrating  Eq.~\eqref{e:confevol}, the Kramers expression gives the dimensionless polymer contribution to the first normal stress difference as \citep{dpl2}
\begin{gather}
\tNIp = 3 f \,\bigl(M_{zz}-M_{rr}\bigr).
\label{e:kramers}
\end{gather}
Then, with the standard definition of the extensional viscosity in uniaxial extension \citep{dpl1}, and using \(\ckBT\,\lamI\) as the characteristic polymer viscosity scale, the dimensionless polymer contribution to the extensional viscosity is
\begin{gather}
\teetap = \frac{\tNIp}{\Wi}.
\label{e:teetapdef}
\end{gather}

The quantities \(\tNIp\) and \(\teetap\) may thus be interpreted as the intrinsic average single-chain contributions to the solution normal stress difference and extensional viscosity. Although solutions with \(\phi<1\) are dilute at equilibrium, significant stretching can enhance hydrodynamic coupling between chains, making the macroscopic behaviour sensitive to concentration \citep{clasenetal, Prabhakar2016, Prabhakar2017-ri}. In this mean-field picture, the average single-chain contribution to stress and viscosity depends on chain density through the concentration dependence of \(\nu\) in Eq.~\eqref{e:confevol}. In the conventional FENE-P model, by contrast, all such concentration dependence is neglected: \(\nu=1\) always, and \(\tNIp\) and \(\teetap\) are independent of \(\phi\).

\subsection{\label{sec:numerical} Numerical solution}

In uniaxial extensional flows starting from equilibrium, only the diagonal components of the conformation tensor are nonzero, and it is convenient to write the evolution equations in strain form. With Hencky strain \(\varepsilon\) as the independent variable,
\begin{equation}
\frac{d M_{ii}}{d\varepsilon}
=
\left(
2\alpha_i - \frac{\theta}{\Wi}
\right) M_{ii}
+
\frac{\sigma}{\Wi},
\qquad i=1,2,3,
\label{eq:ext-strain-ode}
\end{equation}
where for uniaxial extension
\begin{gather}
(\alpha_1,\alpha_2,\alpha_3)=\left(-\tfrac{1}{2},-\tfrac{1}{2},1\right),
\end{gather}
\begin{gather}
M = M_{11}+M_{22}+M_{33},
\end{gather}
and
\begin{gather}
\theta(M)=\frac{f(M)}{\nu(M)},
\qquad
\sigma(M)=\frac{1}{3\,\nu(M)}.
\end{gather}
Here \(f(M)\) is given by Eq.~\eqref{e:FEf}. In the C2D2 model, \(\nu(M;\phi, \NK,\hsK)\) is computed using the blob model of Prabhakar and co-workers \citep{Prabhakar2016, Prabhakar2017-ri}; in the FENE-P model, \(\nu=1\).
For efficiency, the functions \(\nu(M)\), \(\theta(M)\), and \(\sigma(M)\) are precomputed on a grid over the range \(M \in [1,\NK]\) and stored as lookup tables, which are then interpolated during time integration. Time is advanced as an auxiliary variable according to
\begin{gather}
\frac{dt}{d\varepsilon} = \frac{\Deo}{\Wi(\varepsilon)},
\label{eq:t-ode}
\end{gather}
with \(\Wi\) evaluated at each strain step from the balance, Eq.~\eqref{e:Wi-visc}, after substituting
\begin{gather}
R = e^{-\varepsilon/2}.
\end{gather}
The system of Eqs.~\eqref{eq:ext-strain-ode} and \eqref{eq:t-ode} is integrated forward in strain from \(t=0\). For a chain that is relaxed at the start of thinning, the initial condition is the equilibrium coil,
\begin{gather}
M_{11}=M_{22}=M_{33}=\frac{1}{3}.
\end{gather}
To represent the partial prestretch that the bridge-formation protocol can leave behind, the code also admits a non-equilibrium start, obtained by deforming the equilibrium coil affinely through an initial Hencky strain \(\eps_0\),
\begin{gather}
M_{ii}(0)=\tfrac{1}{3}\,e^{2\alpha_i\eps_0},
\label{e:ic-prestretch}
\end{gather}
so that, for uniaxial extension, \(M_{zz}(0)=\tfrac{1}{3}e^{2\eps_0}\) and \(M_{rr}(0)=\tfrac{1}{3}e^{-\eps_0}\), the equilibrium start being recovered at \(\eps_0=0\). The integration terminates either at a prescribed maximum strain \(\varepsilon_{\max}=25\) or when the chains approach full extension, \(M=0.95\,\NK\), whichever occurs first.

This forward march uses Hencky strain as the independent variable only while the strain rate stays appreciable. The strain-form equations carry factors of \(1/\Wi\), including the time map \(dt/d\varepsilon=\Deo/\Wi\) of Eq.~\eqref{eq:t-ode}, that stiffen when \(\Wi\) is small. This happens at low Deborah number: the initial Weissenberg number \(\Wio=\Hv\Deo\) is then itself small, and the early thinning proceeds at low \(\Wi\) until the strain rate  rises as the neck thins. The integrator therefore switches automatically from strain to dimensionless time whenever \(\Wi\) falls below a small threshold, marching in time through these low-rate intervals and reverting to strain once \(\Wi\) recovers.

These computations use an in-house code, written in Julia, for steady and unsteady homogeneous extensional flows \citep{c2d2jl, c2d2report}. The neck Weissenberg number \(\Wi\) is evaluated at each instant from the balance, Eq.~\eqref{e:Wi-visc}. Although the code is multimode, all results shown in the present paper are for the single-mode case. As shown by \citet{entovhinch}, the contributions of  higher relaxation modes matter little in CBR since the selected \(\Wi\) stays of order unity: the faster modes are never strongly actuated, and a multi-mode treatment leaves the results essentially unchanged. The transient trajectories are advanced with a second-order predictor-corrector exponential-time-differencing scheme with adaptive stepping. The steady-state response at fixed \(\Wi\), obtained by setting the left-hand sides of Eq.~\eqref{eq:ext-strain-ode} to zero and solving for the \(M_{ii}\), is computed separately using an adaptive pseudo-arclength continuation method with a Newton-Armijo predictor-corrector solver.

A single run is specified by the choice of drag model -- constant drag, which recovers the FENE-P model with \(\nu=1\), or the full C2D2 closure -- together with the dimensionless inputs: the Deborah number \(\Deo\), the concentration \(\phi\), the finite-extensibility parameter \(\NK\), and the hydrodynamic-interaction parameter \(\hsK\). The prefactor \(\UI\) is pre-computed from these and retains a weak dependence on \(\hsK\) and \(\NK\) through the equilibrium draining ratio even in the constant-drag model, approaching \(0.325\) only as \(\NK\to\infty\). The primary outputs are the transient trajectories \(M(\varepsilon)\), \(\Wi(\varepsilon)\), and \(\teetap(\varepsilon)\), from which we post-process the two quantities used throughout the paper: the Weissenberg number at the elastic point \(\Wie\) and the neck radius at the elastic onset \(\Rve\), both introduced below. Results are obtained by sweeping over the Deborah number at a fixed drag model, \(\NK\), \(\hsK\), and \(\phi\). The steady-state response (which does not depend on \(\Deo\)) is computed once per sweep.

\begin{figure*}[t]
    \centering
    \includegraphics[width=\textwidth]{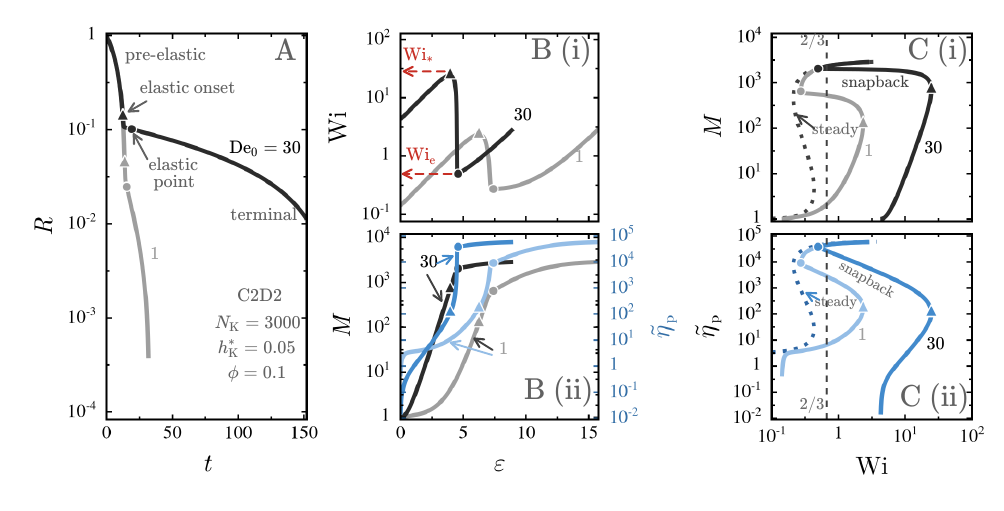}
    \caption{
    Anatomy of a capillary-thinning trajectory and definition of the elastic Weissenberg number \(\Wie\). Predictions of the C2D2 model for a single-mode liquid bridge at fixed \(\NK=3000\), \(\hsK=0.05\), and \(\phi=0.1\), shown for two Deborah numbers, \(\Deo=1\) (light) and \(\Deo=30\) (dark).
    (A) Decay of the normalized neck radius \(R(t)\), tracing the succession of stages: an initial pre-elastic (viscous--capillary) thinning, the elastic onset, the elastic point, the elastic stage, and the terminal stage of finite-extensibility-limited thinning.
    (B) Evolution with Hencky strain \(\varepsilon=-2\ln R\): (i) the instantaneous Weissenberg number \(\Wi(\varepsilon)\) at the neck, whose local maximum marks the elastic onset \(\Wive\) and whose subsequent minimum defines the elastic point \(\Wie\); and (ii) the chain stretch \(M(\varepsilon)\) (left axis) together with the transient polymer extensional viscosity \(\teetap(\varepsilon)\) (right axis).
    (C) The same transients projected onto the phase-space planes (i) \((M,\Wi)\) and (ii) \((\teetap,\Wi)\), each shown with the steady-state response at the same concentration (dotted); the retrograde arm between the elastic onset and the elastic point, along which \(\Wi\) drops sharply at nearly constant stretch, is referred to as \emph{snapback}.
    Throughout, triangular markers identify the elastic onset and circular markers the elastic point; the dashed line marks the Entov--Hinch value \(\Wie=2/3\) (vertical in C).
    }
    \label{fig:intro}
\end{figure*}

\section{Results}
\label{sec:results}
We now turn to the model's predictions, working throughout with the elastic point \(\Wie\) and the elastocapillary Pipkin diagram introduced above. We first establish the qualitative anatomy of a capillary-thinning trajectory and make the definition of \(\Wie\) precise (Sec.~\ref{sec:qualitative}), then chart the Pipkin diagram (Sec.~\ref{sec:pipkin}) and trace how the underlying thinning transients depend on concentration, Deborah number, and molecular parameters, each bounded by the steady-state response, showing in particular how the coil-stretch hysteresis of the C2D2 model can drive the selected \(\Wie\) below the Entov--Hinch prediction of 2/3 (Sec.~\ref{sec:transients}). We then examine the effect of chains left prestretched at the start of thinning (Sec.~\ref{sec:prestretch}). A scaling analysis of the stress balance (Sec.~\ref{sec:pipkin_master_plots}) then rationalizes the operating curves on the Pipkin diagram -- the dilute plateau, the corner, and the finite-extensibility-driven post-corner rise -- and underpins the renormalization that collapses them onto a common master curve (Sec.~\ref{sec:master}). Finally, we compare these predictions with experimental CBR data across seven datasets (Sec.~\ref{sec:exptl}).

\subsection{Qualitative features of capillary thinning}
\label{sec:qualitative}

Figure~\ref{fig:intro} presents representative capillary-thinning trajectories: the curves are single-mode C2D2 model predictions at fixed \(\NK=3000\), \(\hsK=0.05\), and \(\phi=0.1\), and for two Deborah numbers, \(\Deo=1\) and \(\Deo=30\). The terminology and definitions that follow are features of the thinning dynamics and do not depend on the choice of constitutive model. Panel A shows the decay of the normalized neck radius \(R(t)\). Panel B follows the evolution with Hencky strain \(\varepsilon=-2\ln R\): (i) the instantaneous Weissenberg number \(\Wi(\varepsilon)\) at the neck, and (ii) the chain stretch \(M(\varepsilon)\) together with the transient polymer extensional viscosity \(\teetap(\varepsilon)\). Panel C projects the same transients onto the \((M,\Wi)\) and \((\teetap,\Wi)\) planes, where they can be compared with the steady-state extensional response as a function of $\Wi$. We shall refer to the transient trajectories such as those in panel C as ``phase-space'' CBR trajectories.

At early times the polymer stresses are too small to influence the dynamics, and capillarity is balanced almost entirely by the viscous solvent stress. The neck thins linearly in time in this self-similar  viscous \emph{pre-elastic} phase, with a strain rate that starts at the initial value \(\Wio=\Hv\Deo\) set by the Deborah number [Eq.~\eqref{e:Wio}] and grows as the neck thins. On the semilogarithmic axes of panel B\,(i), \(\Wi\) rises almost linearly with the Hencky strain \(\varepsilon\) through this stage, indicating that \(\Wi\) grows as a power law in the decreasing neck radius \(R\) (equivalently, exponentially in \(\varepsilon\)). As \(\Wi\) climbs, the chains begin to stretch.

The two Deborah numbers differ markedly in when this stretching sets in. For \(\Deo=30\) the initial Weissenberg number \(\Wio=\Hv\Deo\approx4.3\) already exceeds the coil-stretch value \(\Wics=1/2\), and panel B\,(ii) shows the stretch \(M\) and the polymer extensional viscosity \(\teetap\) growing from the very start of thinning. For \(\Deo=1\), by contrast, \(\Wio\approx0.14\) lies well below \(\Wics\): the chains stay essentially coiled and the polymer stresses negligible through the early viscous regime, and appreciable stretching is delayed until the rising \(\Wi\) crosses \(\Wics\). The phase-space trajectories in panels C\,(i) and C\,(ii) show this directly: the \(\Deo=1\) trajectory hugs the low-stretch, low-viscosity corner of the diagram and lifts off only once \(\Wi\) exceeds \(\Wics=1/2\), whereas rapid stretching begins in the \(\Deo=30\) trajectory right from the start.

Once the polymer stresses become comparable to capillarity, the system enters the \textit{elastic} regime, in which the bridge effectively selects its own thinning rate. In panel B\,(i), the entry into this regime is marked by a sharp drop in \(\Wi\). The local maximum attained in \(\Wi\) thus identifies the \textit{elastic onset}, \(\Wive\). During the elastic stage, \(\Wi\) subsequently attains a minimum. We define this minimum as the \textit{elastic point}, and the value of \(\Wi\) there as the \textit{elastic Weissenberg number},
\begin{gather}
\Wie = \lamI\,\dot{\varepsilon}_{\mathrm e}.
\end{gather}
Equivalently, \(\Wie\) is associated with the slowest local exponential decay rate of \(R(t)\) in the elastic regime, that is, with the smallest slope of a tangent drawn on a semilogarithmic plot of \(R\) against \(t\). In Fig.~\ref{fig:intro}, and in all subsequent trajectory plots, triangle and circle symbols mark the elastic onset and the elastic point, respectively, on each trajectory.

The phase-space projections in panel C bring out the role of the steady state. The dotted curves are the steady-state stretch and extensional viscosity as functions of \(\Wi\); because the coil-stretch response is hysteretic, each is multivalued, folding back between a weakly stretched (coil) branch and a highly stretched branch. A capillary-thinning trajectory does not trace this steady curve. Instead the flow drives it to higher \(\Wi\), and it is only near the elastic point that the transient collides with the steady response. How far the trajectory is pushed before this collision sets the selected \(\Wie\): for \(\Deo=1\) the elastic point lies close to the low-\(\Wi\) steady branch, whereas for \(\Deo=30\) the trajectory is carried to substantially higher \(\Wi\) before it turns. The dashed line in panels C\,(i) and C\,(ii) marks the Entov--Hinch value \(\Wie=2/3\) \citep{entovhinch}: the elastic points of these C2D2 trajectories lie to its left, showing that the conformation- and concentration-dependent drag can select an elastic Weissenberg number \emph{below} \(2/3\), a result we develop in Sec.~\ref{sec:transients}. Tracking how these collisions systematically migrate the elastic points along the hysteretic steady curve as \(\Deo\), concentration, and molecular parameters are varied is, as the following sections show, the key to understanding the dependence of \(\Wie\) on geometry and material properties.

\begin{figure*}[t!]
    \centering
    \includegraphics[width=\textwidth]{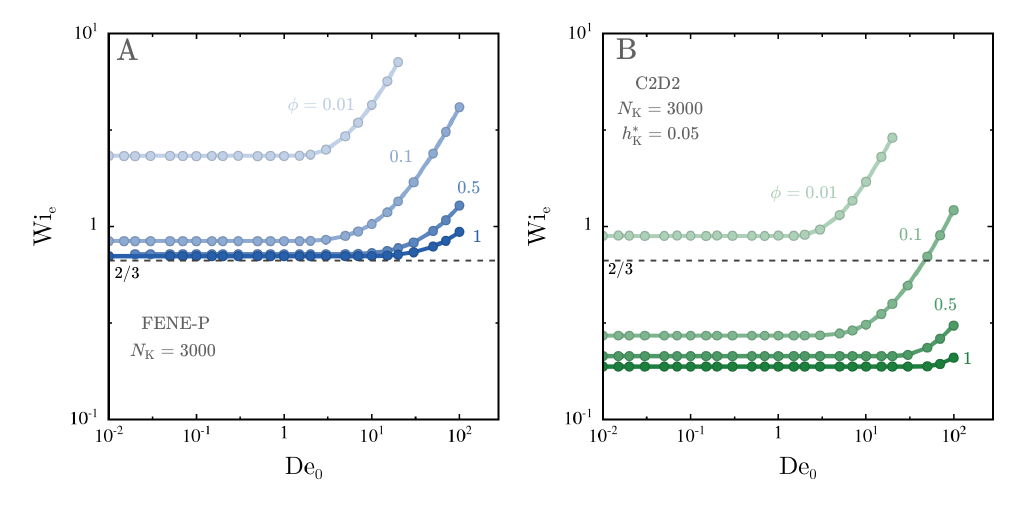}
    \caption{
    Elastocapillary Pipkin diagrams: the self-selected elastic Weissenberg number \(\Wie\) plotted against the Deborah number \(\Deo\) at fixed \(\NK=3000\) and \(\hsK=0.05\), for several concentrations \(\phi=c/\cstar\) (\(\phi=0.01,\,0.1,\,0.5,\,1\)). (A) The conventional FENE-P model; (B) the C2D2 model with conformation- and concentration-dependent drag. Each curve shows a low-\(\Deo\) plateau followed by a post-corner rise; the dashed horizontal line marks the Entov--Hinch value \(\Wie=2/3\). In the FENE-P model the plateaus approach \(2/3\) from above as \(\phi\) increases, whereas in the C2D2 model, the plateaus are driven below \(2/3\) at the higher concentrations.
    }
    \label{fig:pipkin_fenep_c2d2}
\end{figure*}

\subsection{The CBR Pipkin diagram}
\label{sec:pipkin}

We now turn to the dependence of the self-selected elastic Weissenberg number on bridge geometry and concentration. As discussed in the Introduction, in the CBR literature, the reciprocal of the strain rate at the elastic point is often reported as an ``apparent relaxation time'' \(\lame\), so that the ratio \(\lame/\lamI = 1/\Wie\). As already suggested by Eq.~\eqref{e:Wi-visc}, however, the transient evolution of \(\Wi\) is not generally free of geometric influence: it depends not only on the polymer parameters, but also on \(\Deo\). We therefore take  \(\Wie\) as the single primary observable extracted from a transient CBR trajectory, and plot it in Fig.~\ref{fig:pipkin_fenep_c2d2} against \(\Deo\) at fixed \(\NK\) and \(\hsK\), with the concentration \(\phi\) varied across the curves. Following the broader viscoelastic-flow literature \citep{Ewoldt2017-sx, Pipkin1986-wg}, we refer to these \(\Wie(\Deo)\) plots as elastocapillary \emph{Pipkin diagrams} and to the individual curves at fixed \(\NK\),  \(\hsK\) and \(\phi\) as \textit{Pipkin curves}.

The most immediately arresting feature of Fig.~\ref{fig:pipkin_fenep_c2d2} is the close similarity of the Pipkin curves, both as \(\phi\) is varied within each model and between the two models in panels A and B. In every case the curve has the same generic shape: a low-\(\Deo\) plateau, on which \(\Wie\) is essentially independent of the bridge geometry, followed by a corner beyond which \(\Wie\) rises with \(\Deo\). Varying the concentration shifts a curve in a systematic way: as \(\phi\) increases, the plateau value falls and the corner moves to larger \(\Deo\), so that the family of curves marches downward and to the right. This shared shape is the key observation underlying the master-plot construction developed later: because all the curves have the same generic form, an appropriate rescaling of \(\Wie\) and \(\Deo\) should collapse them onto a single master curve (Sec.~\ref{sec:master}).

The principal difference between the two models is where these plateaus sit relative to the Entov--Hinch value \(\Wie=2/3\) \citep{entovhinch}. In the conventional FENE-P model (panel A) the plateau decreases with concentration but approaches \(2/3\) from above, never dropping below it. In the C2D2 model (panel B), by contrast, the conformation- and concentration-dependent drag pushes the plateaus lower still, and at the higher concentrations they fall below \(2/3\). This below-\(2/3\) selection, which the constant-drag FENE-P model cannot produce, is the central qualitative signature of the C2D2 physics, whose origin we trace in Sec.~\ref{sec:transients}.

The post-corner rise of \(\Wie\) in each Pipkin curve is closely related to the observations of \citet{Gaillard2024-xu}, who showed experimentally that the apparent relaxation time measured for the same polymer solution varied systematically with the end-plate radius of the CBR device, first increasing strongly at small radii and then beginning to saturate at larger radii. Since \(\Wie \sim 1/\lame\) and \(\Deo \sim 1/R_0\), the Pipkin diagram in Fig.~\ref{fig:pipkin_fenep_c2d2} is, in effect, a diagonally flipped representation of a plot of \(\lame\) against the initial neck radius \(R_0\), and is therefore qualitatively consistent with the trends observed by \citet{Gaillard2024-xu}. We shall later present the data of \citeauthor{Gaillard2024-xu} on a Pipkin diagram, and compare predictions of the C2D2 model.
We now turn to the transient trajectories in phase space that underlie these Pipkin curves and provide the physical basis for the scaling picture.

\begin{figure*}[t]
    \centering
    \includegraphics[width=\textwidth]{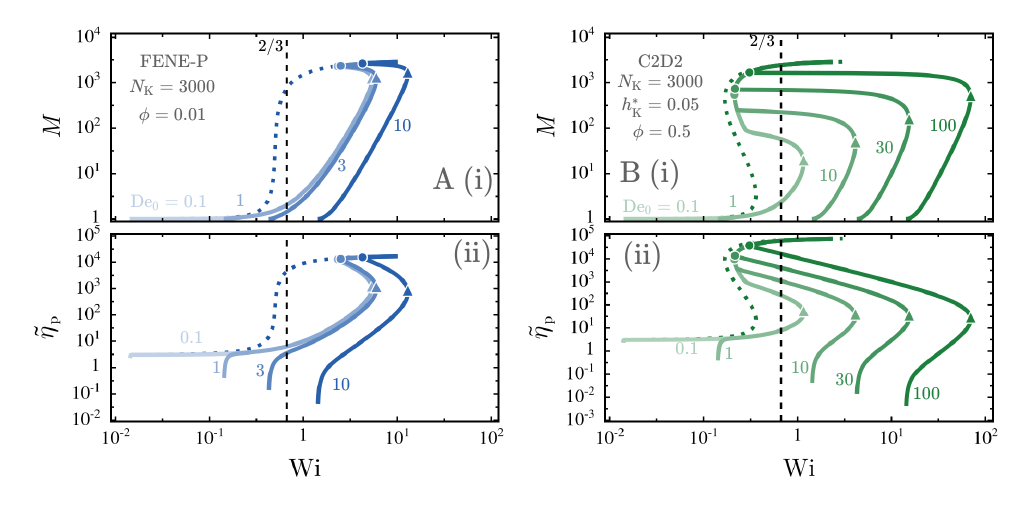}
    \caption{
    Capillary-thinning transients in phase space for the FENE-P and C2D2 models. (A) FENE-P at \(\phi=0.01\) and (B) C2D2 at \(\phi=0.5\), both at \(\NK=3000\) and \(\hsK=0.05\). Within each, (i) plots the chain stretch \(M\) and (ii) the transient polymer extensional viscosity \(\teetap\) against the instantaneous Weissenberg number \(\Wi\), for the Deborah numbers \(\Deo\) indicated alongside the curves. Solid curves are the transient trajectories; the dotted curve is the steady-state response at the same concentration. Triangles mark the elastic onset and circles the elastic point; the corresponding \(\Wie\) values appear on the Pipkin curves of Fig.~\ref{fig:pipkin_fenep_c2d2}. The dashed vertical line marks the Entov--Hinch value \(\Wie=2/3\).
    }
    \label{fig:phase_space}
\end{figure*}

\subsection{Parameter dependence of capillary-thinning transients}
\label{sec:transients}

Figure~\ref{fig:phase_space} traces capillary-thinning transients through phase space, contrasting the FENE-P model at the highly dilute \(\phi=0.01\) (panel A) with the C2D2 model at the moderately dilute \(\phi=0.5\) (panel B). Within each panel, (i) and (ii) project the trajectories onto the \((M,\Wi)\) and \((\teetap,\Wi)\) planes, with the steady-state response shown dotted and the curves labelled by \(\Deo\). The triangular markers identify the elastic onset and the  circular  markers are the elastic points  whose \(\Wie\) values are those plotted in Fig.~\ref{fig:pipkin_fenep_c2d2}. In the FENE-P model the steady-state (dotted) curves in panel A rise through the coil-stretch transition at the critical value \(\Wics=1/2\), whereas in panel B,the steady-state curves of the C2D2 model are hysteretic, folding back between the lower weakly stretched (coiled)  and the higher  highly-stretched branches.

As already suggested by \citet{Prabhakar2016}, capillary-thinning trajectories viewed in \((\Wi,M)\) and \((\Wi,\teetap)\) phase space exhibit characteristic features that are central to understanding how the elastic point varies with \(\phi\) and \(\Deo\). Since \(\Wio=\Hv\Deo\), increasing \(\Deo\) at fixed concentration shifts the initial starting point of each trajectory rightward. Once chains begin to stretch affinely in the viscous regime,  both \(M\) and \(\teetap\) grow strongly as power laws of \(\Wi\); following the elastic onset, \(\Wi\) drops rapidly while \(M\) -- and hence the polymer stress -- stays nearly constant, so that \(\teetap\simeq M/\Wi\) (Eqs.~\eqref{e:kramers} and~\eqref{e:teetapdef}) increases as \(\Wi^{-1}\) along the retrograde arms of the trajectories. We refer to these retrograde branches as \textit{snapback} segments. In every case -- at any \(\phi\) and \(\Deo\), and for both the FENE-P and C2D2 models -- the steady-state curve acts as an attractor in the elastic regime: the snapback carries the trajectory back toward the steady response, which it approaches but never crosses. A crossing would require the left-hand side of the conformation evolution equation~\eqref{e:confevol} to vanish -- the steady-state condition itself -- so each trajectory instead tracks the steady curve from just to its right, at marginally higher \(\Wi\). The elastic point is selected along this near-coincident stretch, and the two regimes that follow differ only in \emph{where} along the steady curve it falls.

Consider first the FENE-P predictions in panel A. At this high dilution the chains reach the elastic onset already close to full stretch, with \(M\) near its maximum value \(\NK\), so the elastic points are limited by the finite-extensibility-saturated branch of the steady-state curve; the snapback segments collide with it at values of \(\Wie\) well above the Entov--Hinch value \(2/3\). We call these \emph{finite-extensibility (FE) limited} elastic points. Increasing \(\Deo\) shifts the trajectories rightward, but only once the initial state lies beyond the coil-stretch threshold: for the smaller values \(\Deo=0.1\) and \(1\), where \(\Wio=\Hv\Deo<\Wics\), the trajectories are practically indistinguishable and share the same elastic point, and hence the same \(\Wie\). This is why the FE-limited Pipkin curves for \(\phi=0.01\) and \(0.1\) in Fig.~\ref{fig:pipkin_fenep_c2d2} display a low-\(\Deo\) plateau; only once
\begin{gather}
\Deo > \frac{\Wics}{\Hv} \simeq 3.5,
\end{gather}
so that \(\Wio>\Wics\), does the FE-limited \(\Wie\) begin to increase approximately linearly with \(\Deo\), reflecting the rightward shift of the trajectories.

At higher concentrations the elastic onset occurs at smaller stretch, so the snapback segments can reach the steady-state curve while it is still rising through the coil-stretch transition. The elastic point is then \emph{stretch-to-coil-transition (SCT) limited}, with \(\Wie=(4/3)\,\Wisc\) (derived in Appendix~\ref{app:scaling-derivation}). In the FENE-P model there is no coil-stretch hysteresis, so \(\Wisc=\Wics=1/2\) and this plateau coincides with the Entov--Hinch value \(2/3\) -- the level approached from above by the FENE-P Pipkin curves for \(\phi=0.5\) and \(1\) in Fig.~\ref{fig:pipkin_fenep_c2d2}.

Panel B shows how this picture changes under the C2D2 model. The conformation- and concentration-dependent drag makes the steady-state curve \textit{hysteretic}, so that the stretch-to-coil transition occurs at a value \(\Wisc\) that lies below \(\Wics=1/2\) and, crucially, decreases with concentration in dilute solutions. The SCT-limited plateau \(\Wie=(4/3)\Wisc\) therefore falls below the Entov--Hinch value \(2/3\), and sinks further as \(\phi\) increases. This is the origin of the below-\(2/3\) plateaus of the C2D2 Pipkin curves in panel~B of Fig.~\ref{fig:pipkin_fenep_c2d2}, and it proves essential for reconciling the experimental data. At sufficiently large \(\Deo\) the rightward shift of the trajectories eventually drives even these elastic points back onto the FE-limited branch, so that \(\Wie\) rises again; because the onset still occurs at small stretch over a finite range of \(\Deo\), the corner of the moderately dilute Pipkin curves shifts to larger \(\Deo\) as \(\phi\) increases.

\begin{figure*}[t]
    \centering
    \includegraphics[width=\textwidth]{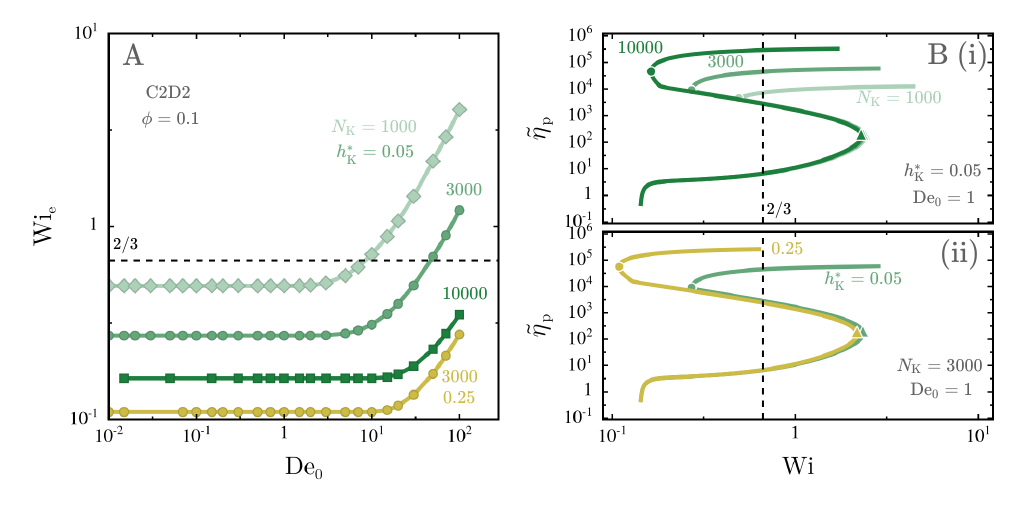}
\caption{
Effect of the finite-extensibility parameter \(\NK\) and the drag parameter \(\hsK\) in the C2D2 model, at fixed concentration \(\phi=0.1\).
(A) Elastocapillary Pipkin curves \(\Wie\) versus \(\Deo\). The upper three curves vary \(\NK=1000,3000,10000\) at fixed \(\hsK=0.05\); the lowest curve raises the drag to \(\hsK=0.25\) at \(\NK=3000\). The dashed line marks the Entov--Hinch value \(\Wie=2/3\).
(B) The corresponding \(\teetap\)-versus-\(\Wi\) phase-space trajectories at \(\Deo=1\): (i) varies \(\NK=1000,3000,10000\) at \(\hsK=0.05\), and (ii) compares \(\hsK=0.05\) and \(0.25\) at \(\NK=3000\). The vertical dashed line marks \(\Wi=2/3\). Because each parameter set has a different steady-state attractor, the steady-state curves are omitted here for clarity. Triangles mark the elastic onset and filled circles the elastic point.
}
\label{fig:hsK_NK_effect}
\end{figure*}

Beyond concentration, the C2D2 predictions depend on the molecular parameters, but again only through a systematic shift of the same generic curves. Figure~\ref{fig:hsK_NK_effect} illustrates this at fixed \(\phi=0.1\). In the Pipkin diagram (panel A), increasing the contour length \(\NK\) lowers the plateau and moves the corner to larger \(\Deo\), and increasing the drag parameter \(\hsK\) lowers the plateau further still. The phase-space trajectories (panel B) follow exactly the rules established in Fig.~\ref{fig:phase_space}: affine stretching, a sharp snapback, and an elastic point (filled circle) selected where the snapback segment collides with the steady-state curve. Here the steady curves themselves are omitted for clarity, since each \((\NK,\hsK)\) pair has its own steady-state attractor, but the elastic points are fixed by those now-invisible collisions exactly as before. These parameters lower the plateau because they reshape the hysteresis loop of the steady curve: neither \(\NK\) nor \(\hsK\) alters the coil--stretch threshold \(\Wics=1/2\), but both deepen the hysteresis and widen its window by lowering the stretch-to-coil value \(\Wisc\), so that the SCT-limited plateau \(\Wie=(4/3)\Wisc\) drops with it. In every case the plateau-and-post-corner shape is preserved.

% 2026-06-28: former \subsection{Effect of conformation- and concentration-dependent drag} (label sec:c2d2)
% folded into Sec.~\ref{sec:transients}; fig:hsK_NK_effect (\NK/h*K dependence) now belongs to that section.

\subsection{Effect of prestretch}
\label{sec:prestretch}
% Aru 2026-06-17: fig:prestretch added (C2D2 model, fine data). NOTE: changed the model reference below from FENE-P to C2D2 to match the figure — confirm this is the intended wording.
The analysis so far assumes the chains are at equilibrium at the commencement of self-similar capillary thinning ($t=0$), when the stress balance first becomes a valid description of the macroscopic dynamics at the necking plane. In practice, the bridge-formation protocol and the initial non-self-similar thinning at the neck can leave the chains partially stretched at \(t=0\). We now examine, within the C2D2 model, how a non-equilibrium initial stretch \(M(0)>1/3\) shifts the transient trajectories and the selected \(\Wie\).

\begin{figure*}[t]
    \centering
    \includegraphics[width=\textwidth]{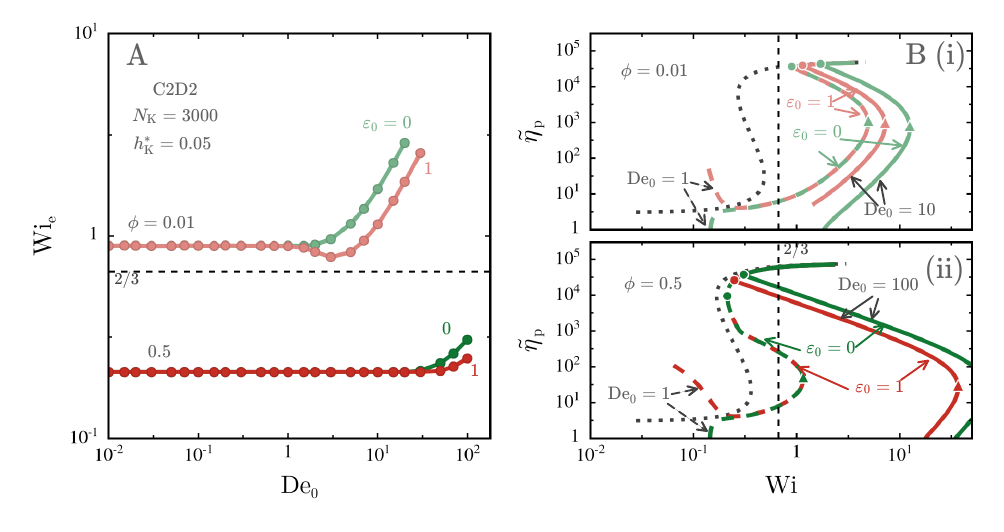}
    \caption{
    Effect of prestretch in the C2D2 model at \((\NK,\hsK)=(3000,0.05)\).
    (A) Elastocapillary Pipkin curves \(\Wie\) versus \(\Deo\) for \(\phi=0.01\) (light shades) and \(\phi=0.5\) (dark shades), each for an equilibrium start (\(\eps_0=0\), green) and a prestretched start (\(\eps_0=1\), red). Prestretch leaves the low-\(\Deo\) plateau essentially unchanged but suppresses the post-corner rise and shifts the corner to larger \(\Deo\); the dashed horizontal line marks the Entov--Hinch value \(\Wie=2/3\).
    (B) The corresponding \(\teetap\)-versus-\(\Wi\) phase-space trajectories at a lower \(\Deo=1\) (dashed) and a higher \(\Deo\) (solid; \(\Deo=10\) in (i), \(\Deo=100\) in (ii)), with the equilibrium start (\(\eps_0=0\)) in green and the prestretched start (\(\eps_0=1\)) in red, as in panel A: (i) \(\phi=0.01\), and (ii) \(\phi=0.5\). At the lower \(\Deo=1\) the prestretched (red) trajectories first relax from their initial stretch before the flow re-stretches them toward the elastic point, so prestretch does not affect the elastic points (circular markers) at low \(\Deo\). At the higher \(\Deo\) prestretch hastens affine stretching so that the trajectories reach the elastic onset (triangular markers) at lower \(\Wi\); where the elastic points for these trajectories are set by the finite-extensibility branch of the steady-state curve, they are displaced to lower \(\Wi\) by the same relative change as the onset points. The dotted curve is the C2D2 steady-state response and the vertical dashed line marks \(\Wi=2/3\).
    }
    \label{fig:prestretch}
\end{figure*}

Figure~\ref{fig:prestretch} contrasts an equilibrium start (\(\eps_0=0\), green) with a prestretched start (\(\eps_0=1\), red) at the highly dilute \(\phi=0.01\) (light shades) and the moderately dilute \(\phi=0.5\) (dark shades). Panel~A shows the effect on the Pipkin curves directly: at low \(\Deo\) the green and red curves are indistinguishable, but beyond the corner the red (prestretched) curve falls below the green (equilibrium) one, so that prestretch \emph{suppresses} the post-corner rise of \(\Wie\) and pushes the corner to larger \(\Deo\). A non-equilibrium stretch imposed at \(t=0\) therefore lowers, rather than raises, the selected elastic Weissenberg number, and does so only where \(\Deo\) is already large enough to lift \(\Wie\) above its plateau.

The phase-space trajectories in panel~B explain why the two ends of the curve respond so differently, the distinction being set by the initial Weissenberg number \(\Wio=\Hv\Deo\) relative to the coil--stretch threshold \(\Wics\). In each sub-panel the dashed trajectories are the lower Deborah number (\(\Deo=1\)) and the solid trajectories the higher (\(\Deo=10\) for \(\phi=0.01\) in (i), \(\Deo=100\) for \(\phi=0.5\) in (ii)). For the dashed \(\Deo=1\) trajectories, \(\Wio<\Wics\): the imposed prestretch leaves the chains extended beyond equilibrium but in a flow too weak to hold them there, so the red (prestretched) trajectory relaxes back toward the coil before the thinning neck re-stretches it through \(\Wics\), rejoining the green (equilibrium) one. From that point the two follow a common affine path to the same elastic onset and the same elastic point (circular markers), which is why the low-\(\Deo\) plateau is untouched. For the solid higher-\(\Deo\) trajectories, by contrast, \(\Wio>\Wics\): the chains stretch affinely from the outset, and the prestretch advances this stretching so that the red trajectory reaches the elastic onset (triangular markers) -- where the polymer stress matches the solvent stress -- at a \emph{lower} \(\Wi\) than the green one. Where the resulting elastic point lies on the finite-extensibility branch of the steady-state curve, it is carried down with the onset by the same relative amount, so \(\Wie\) falls. The suppressed post-corner rise in panel~A is the accumulation of this downward shift across the strongly stretched range of \(\Deo\).

Prestretch does not, however, change the selection mechanism itself: the elastic point is still chosen where the snapback segment meets the steady-state curve, on either its coil-stretch-transition or its finite-extensibility branch. Prestretch alters only the mapping between the geometric Deborah number \(\Deo\) and the state the trajectory has reached by the elastic onset. This points to a natural way to absorb its effect. If the Deborah number is referred not to the initial radius but to the radius \(\Rve\) at the elastic onset -- equivalently, to the onset Weissenberg number through \(\Deve\equiv\Deo/\Rve=\Wive/\Hv\) -- then the prestretched and equilibrium-start data trace the \emph{same} plateau-and-post-corner curve, prestretch merely relocating where a given \(\Deo\) falls along it, and the corner \(\Devebar\) is left unchanged. We develop this renormalization, and the collapse of prestretched and equilibrium-start data onto a single master representation that it makes possible, in the scaling analysis of Sec.~\ref{sec:pipkin_master_plots} (the prestretch case specifically in Sec.~\ref{sec:prestretch-scaling}).

\subsection{Pipkin master plots}
\label{sec:pipkin_master_plots}

\subsubsection{Generic features of the Pipkin curves and trajectories}

\begin{figure*}[t!]
    \centering
    \includegraphics[width=\textwidth]{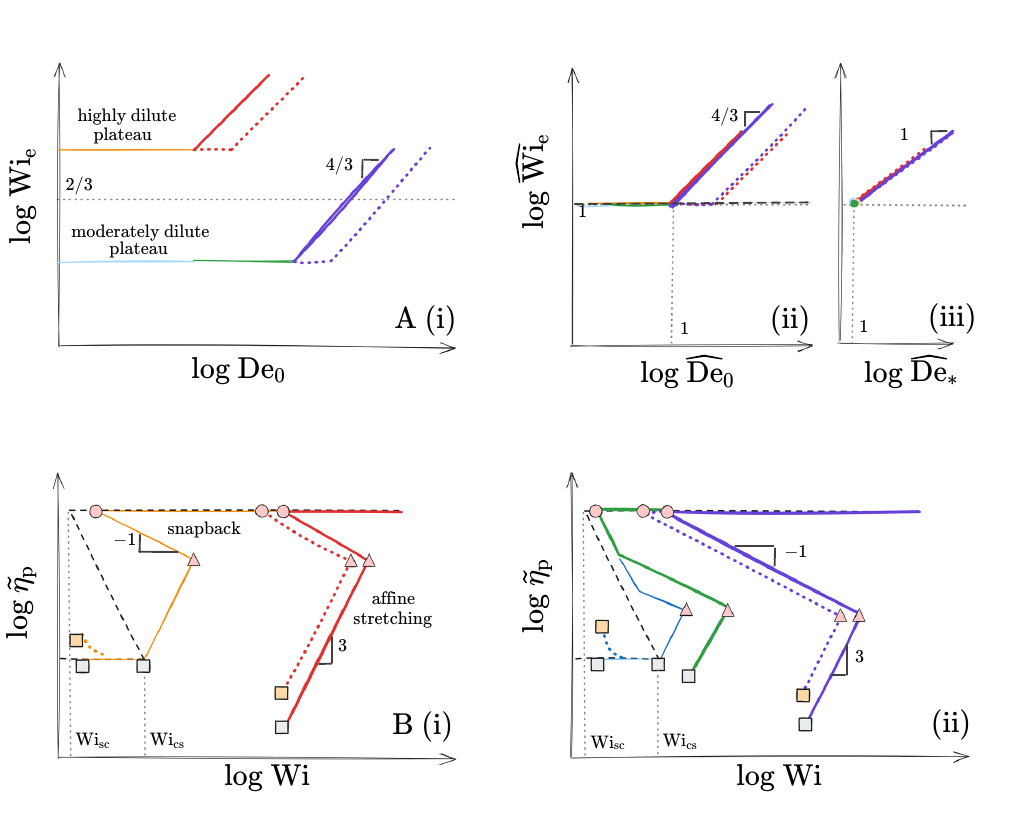}
    \caption{
    Schematic sketches for the scaling analysis. The top row shows the Pipkin diagrams, and the bottom row the underlying transient trajectories in the \((\teetap,\Wi)\) phase-space.
    A\,(i) (top left): the raw Pipkin curves, \(\log\Wie\) versus \(\log\Deo\), with the \emph{highly dilute} plateau (upper) and the \emph{moderately dilute} plateau (lower), the reference level \(\Wie=2/3\), and the post-corner rise of slope \(4/3\).
    (ii) and (iii) (top right): the renormalized Pipkin curves -- \(\log\widehat{\Wi}_\mathrm{e}\) versus \(\log\widehat{\mathrm{De}}_0\) in (ii), with unit-level plateau and slope-\(4/3\) rise, and versus \(\log\widehat{\mathrm{De}}_\ast\) in (iii), collapsed onto the slope-\(1\) master curve.
    B\,(i) (bottom left) and (ii) (bottom right): the corresponding transient trajectories, \(\log\teetap\) versus \(\log\Wi\), for the highly dilute family (i) and the moderately dilute family (ii); each shows the affine-stretching segment of slope \(3\), the snapback segment of slope \(-1\), and the coil--stretch transition Weissenberg numbers \(\Wisc\) and \(\Wics\).
    In either panel, solid curves are cases with no prestretch, while dotted curves have a fixed prestretch. On the phase-space trajectories in the B panels, triangular markers denote the elastic onset and circular markers the elastic point. Gray squares are starting points with zero prestretch, while tan (orange) squares are starting points with prestretch.
    }
    \label{fig:scaling_analysis}
\end{figure*}

Taken together, the qualitative results of the preceding subsections reveal a striking regularity, idealized in the schematic of Fig.~\ref{fig:scaling_analysis}. Across two decades of concentration, for both the FENE-P and C2D2 models, and whether the chains start from equilibrium or prestretched, every Pipkin curve in Figs.~\ref{fig:pipkin_fenep_c2d2}, \ref{fig:hsK_NK_effect}, and~\ref{fig:prestretch} shares the \emph{same} generic shape, sketched in Fig.~\ref{fig:scaling_analysis}A\,(i): a low-\(\Deo\) plateau of \(\Wie\) at a value \(\Wiebar\), followed by a post-corner rise of slope \(4/3\) that sets in at a corner \(\Deo=\Debaro\). Within the dilute regime \(0<\phi\le1\) the curves fall into two families, separated to a first approximation by the Entov--Hinch level \(\Wie=2/3\): a \emph{highly dilute} family with \(\Wiebar>2/3\) and a \emph{moderately dilute} family with \(\Wiebar\lesssim2/3\), the two meeting at a boundary concentration \(\phibar\). The families also differ in their corner: \(\Debaro\) is independent of \(\phi\) in the highly dilute family, but shifts to larger \(\Deo\) with increasing \(\phi\) in the moderately dilute one.

The phase-space trajectories of Figs.~\ref{fig:phase_space} and~\ref{fig:prestretch}, idealized in Fig.~\ref{fig:scaling_analysis}B, explain why the shape is so robust. The location of the elastic point depends on where the snapback segment (slope \(-1\)) following the elastic onset meets the steady-state curve. The elastic onset is, in turn, at the end of the affine-stretching segment (slope \(3\)). A \(\Deo\)-independent elastic point -- and hence the plateau -- arises in two ways. For the initially-weak trajectories (\(\Wio<\Wics\); orange in panel~B\,(i), green in B\,(ii)) the chains creep along the zero-rate viscosity plateau \(\teetapo\) to \(\Wics\) before stretching, following a common path to a common elastic onset, and so to a common elastic point, whatever \(\Deo\). In the moderately dilute family the plateau reaches further, taking in the initially-strong blue trajectory (\(\Wio>\Wics\)): its onset does now shift with \(\Deo\), but the snapback still meets the steady-state curve on its retrograde arm, where the SCT-limited elastic point is pinned at \(\Wie=(4/3)\Wisc\), again independent of \(\Deo\). The post-corner rise, in contrast, collects the trajectories for which increasing \(\Deo\) shifts the starting point rightward until the snapback collides with the finite-extensibility branch, dragging the elastic point upward as \(\Wie\sim\Deo^{4/3}\). This scaling is a simple consequence of the geometry of the power laws governing the affine-stretching and snapback segments of the phase-space trajectories. Only \emph{where} the plateau and corner sit changes from curve to curve. To make the connection explicit, each coloured segment of a Pipkin curve in panel~A of Fig.~\ref{fig:scaling_analysis} is drawn in the colour of the panel-B trajectory that selects its elastic points, so that the plateau and post-corner branches of a given curve carry different colours.

The plateau value \(\Wiebar\) itself is limited differently in the two families. In the highly dilute regime it is set by finite extensibility, lies above the Entov--Hinch value \(2/3\), and decreases as the concentration rises toward \(\phibar\). Once the concentration is large enough (\(\phi\ge\phibar\)) the plateau is instead SCT-limited, \(\Wiebar=(4/3)\Wisc\), and it is here that the coil--stretch transition controls the concentration dependence: in the constant-drag FENE-P model, which has no coil--stretch hysteresis, \(\Wisc=\Wics=1/2\) and the plateau saturates at \(2/3\), whereas in the C2D2 model the conformation- and concentration-dependent drag renders the steady curve hysteretic, so that \(\Wisc\) -- and with it \(\Wiebar\) -- drops below \(2/3\) and falls further with concentration \citep{Prabhakar2016}. Beyond this, the molecular parameters \(\NK\) and \(\hsK\) only relocate a curve without changing its shape, and prestretch merely shifts the corner \(\Debaro\).

Because every Pipkin curve shares this shape, an appropriate rescaling of \(\Wie\) and \(\Deo\) should collapse the whole family onto a single master curve. The natural normalizing factors are precisely the plateau value \(\Wiebar\) and the corner \(\Debaro\), whose origins we have just traced qualitatively; what remains is to estimate them quantitatively. The scaling analysis that provides these estimates extends the classical argument of \citet{entovhinch}; its details are given in Appendix~\ref{app:scaling-derivation}, and the resulting expressions for \(\Wiebar\), \(\Debaro\), and the boundary concentration \(\phibar\) are summarized next.

\subsubsection{Estimates of plateau and corner values from scaling analysis}

These results are collected in Table~\ref{tab:pipkin_summary} for the two dilute-solution Pipkin-curve families. The boundary concentration \(\phibar\) separating the two regimes is given by
\begin{gather}
\phibar
=
\frac{3}{\teetapo\,\UI\,(\FE \NK)^{3/4}}.
\label{e:phibar}
\end{gather}
Here \(\teetapo\) is the zero-extension-rate plateau value of the dimensionless polymer extensional viscosity, which a perturbation analysis at small Weissenberg number gives as \(\teetapo=3\); and \(\FE\) is an empirical parameter marking the onset of finite-extensibility-dominated behaviour, where the chain stretch reaches \(M\simeq\FE\NK\), which we estimate to be \(\FE\approx0.22\). For \(\phi<\phibar\), the plateau is FE-limited, whereas for \(\phi\ge\phibar\) it is SCT-limited.

\begin{table}[h!]
\centering
\caption{
Summary of the plateau value \(\Wiebar\) and corner Deborah number \(\Debaro\) for the two dilute-solution Pipkin-curve families. The boundary concentration \(\phibar\) separating these regimes is given by Eq.~\eqref{e:phibar}.
}
\label{tab:pipkin_summary}
\begin{tabular}{lcc}
\hline
Concentration regime & \(\Wiebar\) & \(\Debaro\) \\
\hline
Highly dilute, \(\phi < \phibar\)
&
\(\displaystyle
\frac{2}{3}\,\frac{\Wisc}{\Wics}
\left[
\frac{3^{4/3}}{\FE \NK\,(\teetapo \UI\phi)^{4/3}}
\right]
\)
&
\(\displaystyle
\frac{\Wics}{\Hv}
\)
\\[5mm]
Moderately dilute, \(\phi \geq \phibar\)
&
\(\displaystyle
\frac{2}{3}\,\frac{\Wisc}{\Wics}
\)
&
\(\displaystyle
\frac{\Wics}{\Hv}\,\frac{\phi}{\phibar}
\)
\\[2mm]
\hline
\end{tabular}
\end{table}

The two families of Pipkin curves collapse onto a common plot when \(\Wie\) is normalized by \(\Wiebar\) and \(\Deo\) by the corner \(\Debaro\). In the scaled variables the post-corner branch is
\begin{gather}
\frac{\Wie}{\Wiebar}
=
\begin{cases}
1, & \Deo\le\Debaro,\\[2mm]
\left(\dfrac{\Deo}{\Debaro}\right)^{4/3}, & \Deo>\Debaro,
\end{cases}
\label{e:pipkin-collapse}
\end{gather}
so that both families share the same finite-extensibility-controlled rise and differ only through \(\Wiebar\) and \(\Debaro\) -- the renormalized Pipkin plot in the top-right panel~(ii) of Fig.~\ref{fig:scaling_analysis}, where every curve is a unit plateau followed by a slope-\(4/3\) rise.

\begin{figure*}[tp]
\centering
\includegraphics[width=\textwidth]{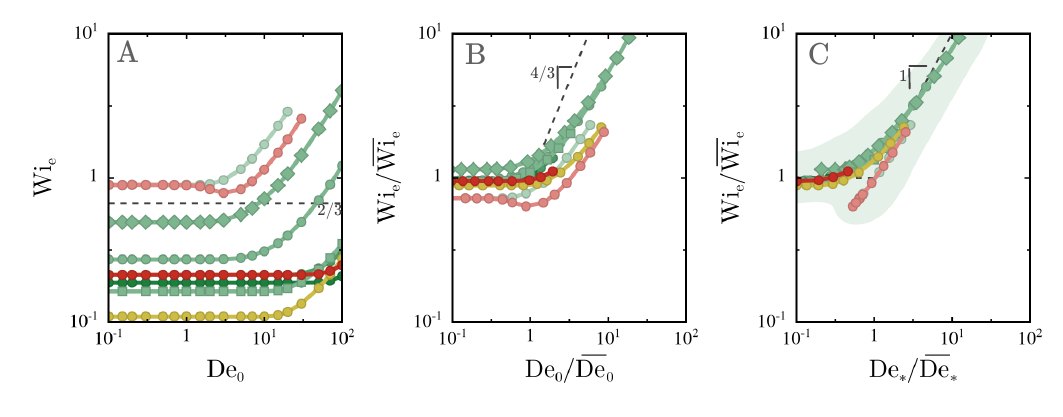}
\caption{
Master collapse of the C2D2 elastocapillary Pipkin curves. (A) Raw Pipkin curves, \(\Wie\) versus \(\Deo\); the dashed line marks the Entov--Hinch value \(\Wie=2/3\). The three panels show all the C2D2 Pipkin curves computed in the preceding sections: the concentration sweep of Fig.~\ref{fig:pipkin_fenep_c2d2}(B) (\(\phi=0.01,0.1,0.5,1\), light to dark green, at \(\NK=3000\), \(\hsK=0.05\)); the finite-extensibility and drag variations of Fig.~\ref{fig:hsK_NK_effect} (\(\NK=1000,3000,10000\) shown as diamonds, circles and squares at \(\phi=0.1\), and the higher drag \(\hsK=0.25\) in yellow); and the prestretched starts of Fig.~\ref{fig:prestretch} (\(\eps_0=1\), red, for \(\phi=0.01\) and \(0.5\)).
(B) The same data with \(\Wie\) scaled by the plateau estimate \(\Wiebar\) and \(\Deo\) by the corner estimate \(\Debaro\) (Table~\ref{tab:pipkin_summary}); the dashed guide has slope \(4/3\).
(C) The collapse onto the slope-\(1\) master line under the onset-referenced abscissa \(\Deve/\Devebar\) [Eq.~\eqref{e:Deve}], which also absorbs the prestretch. The corner here is \(\Devebar=\Kstar\Devebarpred\): the scaling prediction~\eqref{e:devebar} carried by a single empirical factor \(\Kstar\) per \((\NK,\hsK)\) family [Eq.~\eqref{e:kstar-main}], and serving the equilibrium and the prestretched curves. The shaded band is a kernel-density estimate, constructed as described in the SI \citep{SI}, of the full simulated ensemble over which \(\Kstar\) is characterized there -- \(17\) families spanning \(\NK=1000\)--\(58{,}400\) and \(\hsK=0.005\)--\(0.25\) -- on which the present curves lie.
}
\label{fig:pipkin_master}
\end{figure*}

\subsubsection{Accounting for prestretch}
\label{sec:prestretch-scaling}
This collapse is incomplete, however: it removes the concentration dependence, but does not eliminate the shift in the corner value caused by prestretch. A non-equilibrium initial stretch \(M(0)>1/3\) relocates where a given \(\Deo\) falls along the curve, so the equilibrium-start (solid) and prestretched (dotted) data in panel~(ii) share the same shape yet remain shifted relative to one another.

Both shortcomings are cured by referring the Deborah number not to the initial radius \(R_0\) but to the radius \(\Rve\) at the elastic onset. Since the local viscocapillary timescale grows linearly with \(R\), the renormalized Deborah number is
\begin{gather}
\Deve \;\equiv\; \frac{\Deo}{\Rve}
\;=\; \frac{\Wive}{\Hv}.
\label{e:Deve}
\end{gather}
The right-hand side follows from Eq.~\eqref{e:Wit-visc}, where the Weissenberg number \emph{at the elastic onset} \(\Wive\) is identified as the peak \(\Wi\) before the system enters the elastic phase. The renormalized Deborah number is the natural abscissa because the elastic point is itself selected at the onset: whatever the concentration or the prestretch, the trajectory's fate after entering the elastic regime is fixed once \(\Wive\) is reached. Two consequences follow. First, since \(\Deve\sim\Wive\sim\Deo^{4/3}\), and by Eq.~\eqref{e:pipkin-collapse} \(\Wie \sim \Deo^{4/3}\) in the post-corner rise, we obtain a linear scaling for that rise:
\begin{gather}
\Wie \;\sim\; \Deve.
\label{e:masterstar}
\end{gather}
This is the slope-\(1\) collapse shown in the top-right panel~(iii) of Fig.~\ref{fig:scaling_analysis}. Second, the same abscissa absorbs the prestretch. In the \emph{initially-weak stretching} regime (\(\Wio<\Wics\)) the chains relax back toward equilibrium before the flow re-stretches them past \(\Wics\), so the onset -- and hence \(\Deve\) -- is reached along the equilibrium path and is prestretch-independent; in the \emph{initially-strong stretching} regime (\(\Wio>\Wics\)) the prestretch does shift the onset, but \(\Rve\) shifts with it, and the elastic point still tracks the onset stretch exactly as without prestretch (\(\Wie\propto M_*\) on the FE-limited branch, \(\Wie=(4/3)\Wisc\) on the SCT-limited plateau). Plotted against \(\Deve\), therefore, prestretched and equilibrium-start data fall on the \emph{same} master curve; the prestretch only changes which \(\Deo\) maps to a given point.

The corner of this master curve, \(\Devebar=\Wive(\Debaro)/\Hv\), follows from the onset Weissenberg number evaluated at the corner: in the highly dilute family from the \(\Deo\)-independent plateau value~\eqref{e:wive-weak} at \(\Debaro=\Wics/\Hv\), and in the moderately dilute family from~\eqref{e:wive-strong} at \(\Debaro=(\Wics/\Hv)(\phi/\phibar)\). Hence, using the scaling analysis in Appendix~\ref{app:scaling-derivation}, we obtain \(\Devebarpred\), our scaling \emph{prediction} of that corner,
\begin{widetext}
\begin{gather}
\Devebarpred =
\begin{cases}
\dfrac{\Wics}{\Hv}\left(\dfrac{3}{\teetapo \UI\phi}\right)^{1/3}
= \dfrac{\Wics}{\Hv}\,(\FE\NK)^{1/4}\left(\dfrac{\phibar}{\phi}\right)^{1/3}, & \phi\le\phibar,\\[4mm]
\dfrac{\Wics}{\Hv}\left(\dfrac{\phi}{\phibar}\right)^{4/3}\!\left(\dfrac{3}{\teetapo \UI\phi}\right)^{1/3}
= (\FE\NK)^{1/4}\,\Debaro, & \phi>\phibar.
\end{cases}
\label{e:devebar}
\end{gather}
\end{widetext}
The two concentration-regime branches meet at \(\Devebarpred=(\Wics/\Hv)(\FE\NK)^{1/4}\) when \(\phi=\phibar\), so the corner is continuous across the boundary concentration, falling as \(\phi^{-1/3}\) in the highly dilute family and rising as \(\phi\) in the moderately dilute one. Being fixed by the onset, \(\Devebar\) is itself prestretch-independent, so this zero-prestretch estimate applies unchanged to prestretched data at any \(\phi\), \(\NK\), and \(\hsK\). In simulations, \(\Deve=\Deo/\Rve = \Wive/\Hv\) is read off each simulated trajectory through its elastic-onset \(\Wive\); the corner, by contrast, is referred to the prediction~\eqref{e:devebar}, which draws on no quantity extracted from any individual trajectory. How the predicted corner \(\Devebarpred\) relates to the corner values \(\Devebar\) estimated directly from the Pipkin curves themselves  is taken up in the following section. 

\begin{figure*}[p]
\centering
\includegraphics[width=\textwidth]{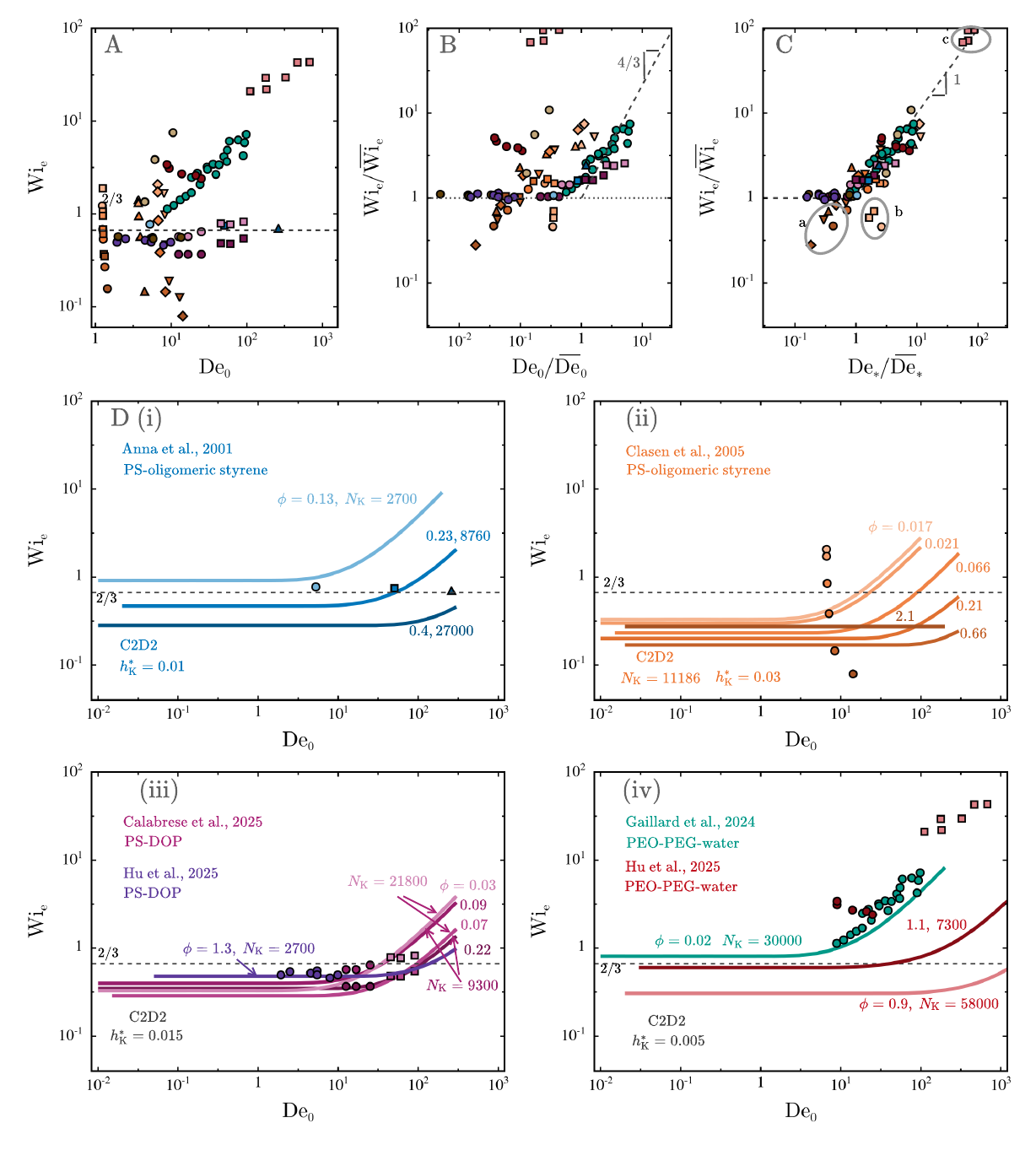}
\caption{
Experimental master Pipkin diagrams with scale factors obtained with the C2D2 model. Each experimental polymer-solvent system is distinguished by a separate  colour:  PS-in-OS system of ~\citet{annamckinley} -- blue \legdot{sysAnna};  PS-in-OS system of ~\citet{clasenetal} -- orange ~\legdot{sysClasen}; PS-in-DOP system of \citet{Calabrese2025-re} -- violet \legdot{sysCalabrese};  PS-in-DOP system of \citet{Hu2025-rt} -- purple \legdot{sysHuPSDOP}; PEO-in-PEG--water system of ~\citet{Gaillard2024-xu} -- green~\legdot{sysGaillard}; PEO-in-PEG--water system of ~\citet{Hu2025-rt} -- crimson~\legdot{sysHuPEO}; PIB--polybutene system of ~\citet{Hu2025-rt} -- brown~\legdot{sysHuPIB}. Within each dataset, lighter to darker shading represents increasing concentration, while marker shapes distinguish molecular weights. Panels A -- C present all the experimental data. (A) Raw elastocapillary Pipkin diagram;  the dashed line marks the Entov--Hinch value \(\Wie=2/3\). (B) Data normalized by the scaling estimates of the plateau \(\Wiebar\) and the corner \(\Debaro\). (C)  Data with the abscissa in terms of the onset-referenced \(\Deve\) normalized by the empirical corner \(\Devebar=\Kstar\Devebarpred\).  Groups \(\mathrm{a}\) and \(\mathrm{b}\) mark outliers that fall well below the predicted master curve, in the moderately dilute and highly dilute regimes respectively; group \(\mathrm{c}\) reaches the master line only through an anomalously small corner shift \(\Kstar\). D(i)--(iv):  Pipkin diagrams showing C2D2 Pipkin curves visually fitted to match experimental data by adjusting a single value $\hsK$ common to the experimental system, the dashed horizontal lines mark \(\Wie=2/3\). In (ii) only the highest-molecular-weight sample data is shown;  the lower molecular-weight samples data and model curves are shown in the SI. The scaling estimates \(\Wiebar\), \(\Debaro\) and \(\Devebarpred\) used in (B) and (C) are computed at the \( \hsK \) values indicated.
}
 \label{fig:exptl}
\end{figure*}

\begin{figure}[tp]
\centering
\includegraphics[width=\columnwidth]{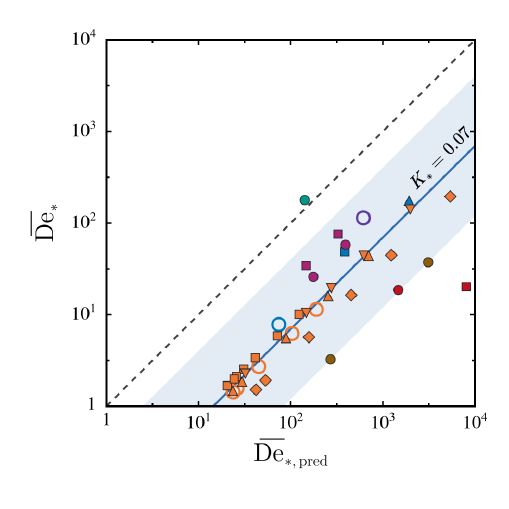}
\caption{
The experimental corner-shift ratio \(\Kstar\), read across the whole body of data --- the measured counterpart of the simulated \(\Kstar\) ensemble characterized in the SI \citep{SI}. Each sample is placed by its scaling-predicted corner \(\Devebarpred\) [Eq.~\eqref{e:devebar}] on the abscissa and by its empirical corner \(\Devebar=\Kstar\Devebarpred\) on the ordinate, so that a system of constant \(\Kstar\) falls along a ray of unit slope; the dashed line marks \(\Kstar=1\), while the solid ray marks the in-band geometric-mean corner shift \(\Kstar\approx0.07\) and the shaded band about it, two geometric standard deviations wide, represents the empirical scatter of the systems that share the common conformation-dependent-drag physics --- the range within which any system governed by that physics alone would be expected to fall; the two systems lying outside it, the \citet{Gaillard2024-xu} PEO solution and the PEO--PEG-8M, are excluded from both estimates. Colours denote the polymer--solvent systems as in Fig.~\ref{fig:exptl}, and marker shapes distinguish molecular weights within a system. Filled symbols carry a \(\Kstar\) measured from the post-corner points of the family; open symbols mark the three low-molecular-weight families that retain no post-corner point of their own, whose \(\Kstar\) is borrowed from the nearest evidence and so lies on a donor's ray by construction (SI \citep{SI}). Most systems gather within this band; the \citet{Gaillard2024-xu} PEO solution sits near \(\Kstar=1\), above it, and the \citet{Hu2025-rt} PEO--PEG-8M solution falls well below it at \(\Kstar\approx0.003\).
}
\label{fig:exptl_kstar}
\end{figure}

\subsubsection{Master plots}
\label{sec:master}

Figure~\ref{fig:pipkin_master} gathers together all the C2D2 Pipkin curves computed in the preceding sections -- the concentration sweep of Fig.~\ref{fig:pipkin_fenep_c2d2}(B), the finite-extensibility and drag variations of Fig.~\ref{fig:hsK_NK_effect}, and the equilibrium-start and prestretched curves of Fig.~\ref{fig:prestretch} -- and asks whether they collapse onto a single master curve. Panel~A shows the raw curves \(\Wie(\Deo)\), whose plateaus span more than a decade in height as \(\phi\), \(\NK\), \(\hsK\), and the prestretch are varied. Panel~B normalizes \(\Wie\) by the plateau value \(\Wiebar\) and \(\Deo\) by the corner \(\Debaro\), bringing the plateaus to a common unit level and the post-corner rises toward the slope-\(4/3\) line. The plateau values \(\Wiebar\) used to collapse the Pipkin curves vertically in panel~B are the scaling estimates of Eq.~\eqref{e:phibar} and Table~\ref{tab:pipkin_summary}; the corner value \(\Debaro\) for each Pipkin curve is calculated, again, from the scaling estimates of Table~\ref{tab:pipkin_summary}. The equilibrium-start curves are brought together well, but it is equally clear that the curves with prestretch \(\varepsilon_0 = 1\) are displaced towards smaller \(\Deo/\Debaro\) and do not collapse onto the same master curve as the curves without prestretch. Normalizing by \(\Debaro\) removes the concentration dependence; it leaves the prestretch shift untouched.

Panel~C accounts for the prestretch as well, by replacing the abscissa with the onset-referenced \(\Deve/\Devebar\) of Eq.~\eqref{e:Deve}. This requires a corner value \(\Devebar\) for each of the Pipkin curves once they are plotted in terms of \(\Deve\), and these we determine empirically, by a two-stage process \citep{SI}. First, we use the scaling prediction \(\Devebarpred\) of Eq.~\eqref{e:devebar}, built from molecular parameters alone. Next, we determine the extra horizontal shift along the logarithmic abscissa, \(\log\Kstar\), that is necessary to bring a whole \(\phi\)-family, for any given set of \((\NK, \hsK, \varepsilon_0)\), onto the unit-slope master line; in other words, \begin{gather}
\Kstar \;\equiv\; \frac{\Devebar}{\Devebarpred},
\label{e:kstar-main}
\end{gather}
which would be unity were the scaling prediction exact. The SI \citep{SI} sets out the protocol for determining \(\Kstar\) and characterizes it over a grid of \(17\) \((\NK,\hsK)\) families spanning the experimental range at five concentrations each, both for \(\varepsilon_0 = 0\) and \(\varepsilon_0 = 1\). The shaded band in panel~C is that ensemble; the curves gathered here lie within it.

We observe that the scaling correction \(\Kstar\) is not universal, and that it is of order unity. It rises monotonically with \(\hsK\), from \(\Kstar\approx0.7\)--\(1.2\) at \(\hsK=0.005\) to \(2.8\)--\(2.9\) at \(\hsK=0.05\), and reaches \(5.8\) at the largest value we have simulated, \(\hsK=0.25\); within a family it drifts by about twofold across concentration, and it is left unchanged by prestretch. At fixed \(\hsK\), however, it changes little across the whole range of \(\NK\). These observations have two important implications. Firstly, that \(\Kstar\) stays of order unity across every \(\phi\), \(\NK\) and \(\hsK\) we have simulated says that the scaling estimate \(\Devebarpred\) captures the universal mechanism that sets the corner. Secondly, that \(\Kstar\) nevertheless varies weakly and systematically with \(\hsK\) and \(\phi\) is the measure of what a scaling argument, by construction, cannot deliver -- the prefactors it discards along the way. It is worth noting that \(\Kstar\) absorbs the discarded prefactors of \emph{both} normalizers, not the corner alone. Because the shift is fixed against the slope-\(1\) post-corner asymptote, on which a horizontal displacement and a vertical one are interchangeable, an error in the predicted corner \(\Devebarpred\) and an error in the predicted plateau \(\Wiebar\) -- the vertical scale of the Pipkin diagram -- enter \(\Kstar\) in exactly the same way and cannot be separated by the collapse; a single \(\Kstar\) stands in for the unknown $O(1)$ prefactors of both.

Set against the wide spread of the raw curves in panel~A -- which fold in the effects not only of concentration and \(\Deo\) but also of prestretch, \(\NK\), and \(\hsK\) -- the quality of the collapse in panel~C is remarkable. A collapse blindly engineered by shifting the predicted Pipkin curves would carry no insight. Obtaining it from the independently predicted factors \(\Wiebar\) and \(\Devebarpred\), with a single shift of order unity per family as the only concession, is what makes the master plot a genuine test of the scaling picture. Our procedure also quantifies, through \(\Kstar\), any extra contributions from non-universal behaviour that a scaling analysis cannot capture.

We take the same approach to the experimental data in Sec.~\ref{sec:exptl}. Deviations of \(\Kstar\) from unity comparable to those found in the simulation results above are to be expected there, and are not by themselves evidence that the model is fundamentally incorrect. The C2D2 model is itself built on scaling arguments, in the form of blob concepts, and thus assumes prefactors of order unity of its own. Knowing the range of \(\Kstar\) this model produces, we can ask of each experimental system whether its \(\Kstar\) is of a magnitude the scaling picture  predicts. If the deviations of \(\Kstar\) from unity are much larger, they point instead to physics that the model does not contain.

\subsection{Experimental master plots}
\label{sec:exptl}
Figure~\ref{fig:exptl} presents the C2D2 predictions alongside experimental CBR data from seven diverse datasets: two different PS-in-oligomeric styrene (OS) Boger fluids \citep{annamckinley, clasenetal}, PS-in-DOP solutions \citep{Calabrese2025-re, Hu2025-rt}, PIB--polybutene Boger fluids \citep{Hu2025-rt},  and two different PEO-in-viscous PEG--water Boger fluids \citep{Gaillard2024-xu, Hu2025-rt}. Together they span three decades in molecular weight and two in concentration, and were measured with a range of capillary-breakup geometries and protocols: step-strain CaBER in the \citeauthor{annamckinley} and \citeauthor{clasenetal} studies, slow-retraction and step-strain CaBER in the \citeauthor{Gaillard2024-xu} and \citeauthor{Calabrese2025-re} studies, and, in the \citeauthor{Hu2025-rt} study, dripping and dripping-onto-substrate across a series of needle sizes, which contributes their PS--DOP, PEO--PEG and PIB--polybutene solutions here. The detailed physical properties and geometric parameters of these systems, including the concentration \(\phi=c/\cstar\), the Kuhn-segment number \(\NK\), the intrinsic relaxation time \(\lamI\), and the geometric quantities needed to form \(\Deo\) and \(\Deve\), are worked out for all samples in the Supplementary Information~\citep{SI}. These studies report measurements of the apparent relaxation time \(\lame\) which are converted here to \(\Wie\) using the sample-wise estimates of \(\lamI\).

The model needs one further input --- the dimensionless HI parameter \(\hsK\). We estimate it under a deliberately tight constraint: a given polymer--solvent system is assigned a \emph{single} \(\hsK\), common to all of its molecular weights and concentrations. We fix \(\hsK\) for each system by raising or lowering it until the family of predicted Pipkin curves -- one per concentration at a given molecular weight -- passes close to the corresponding experimental points, judged \emph{visually} on the Pipkin diagram. The values of $\hsK$ shown in panels~D(i)--(iv) of Fig.~\ref{fig:exptl} were selected after comparisons of model predictions obtained on a coarse grid of $\hsK$ values. With this approach, the agreement is qualitative. Alternatively, we can more precisely fit the \( \hsK \) value to the model Pipkin curve through each dataset of a given molecular-weight and concentration, and then report the mean $\hsK$ value with an estimate of uncertainty.  As also noted by \citet{Prabhakar2016}, the values of \(\hsK\) that are required to describe the data well are much smaller than the fixed-point value of around \( 0.2 \) required for fast convergence to the universal non-draining limit in equilibrium coils \citep{Kroger2000}.

Pooling all seven datasets, panel~A plots the raw \(\Wie\) against \(\Deo\). The points show no apparent common organization. Panel~B rescales every point by the C2D2 estimates of its plateau \(\Wiebar\) and corner \(\Debaro\), obtained using the $\hsK$ values shown in panel D. This brings the plateaus to a common level near unity but collapses the post-corner data only partially. The reason lies in the model's starting assumption: the model predictions and scaling analyses take for granted that self-similar thinning where the simple stress balance is valid begins at $t = 0$, \emph{and} that chains are  at equilibrium at this \(t=0\). In reality, however, the self-similar thinning regime is only entered after the neck becomes sufficiently narrow, and by that time, polymer chains may already be significantly stretched. It is difficult to determine this point and the degree of microscopic affine prestretch \( \varepsilon_0 \) at that point. The \(\Deo\) we assign from the reported initial radius \(R_0\) therefore does not correspond to the true \(t=0\) self-similar-regime onset in the experiment, and the resulting horizontal misregistration is what blurs the collapse in panel~B.

Panel~C removes this ambiguity. Referred to the radius \(\Rve\) at the elastic onset rather than to \(R_0\), the renormalized Deborah number \(\Deve=\lamI\gamma/(\etas\Rve)\) is a purely experimental quantity, independent of the prestretch and of the initial radius \(R_0\) (and hence of \(\Deo\)) (Sec.~\ref{sec:pipkin_master_plots}); \(\Rve\) -- unlike the true \(t=0\) state -- is experimentally well-defined and measured. Each datum can therefore be placed on the \(\Deve\) axis reliably. We then place each dataset on the master plot by the same two-stage construction used for the simulations in Fig.~\ref{fig:pipkin_master}(C). The predicted corner \(\Devebarpred\) follows from Eq.~\eqref{e:devebar} evaluated at each sample's \((\NK,\hsK,\phi)\); being referred to the elastic onset rather than to \(\Deo\), it is prestretch-independent and needs no knowledge of the unmeasured \(t=0\) state, so -- unlike in the previous, \(\Deo\)-anchored normalization -- it \emph{can} be computed for the experiments. The corner-shift ratio \(\Kstar=\Devebar/\Devebarpred\) is then the single horizontal shift, common to all concentrations of a \(\phi\)-family, that carries its post-corner points onto the unit-slope line, determined exactly as for the simulations [Eq.~\eqref{e:kstar-main}]. For the few families that retain no point above the fit cut-off, \(\Kstar\) is borrowed from the nearest available evidence rather than fitted (SI \citep{SI}). With \(\Devebar=\Kstar\Devebarpred\), the data fall onto the slope-\(1\) master line of panel~C.

Whilst there is residual scatter about the predicted master curve in Fig.~\ref{fig:exptl} C, the significance of the data collapse on the predicted master curve is best judged against the scatter in panel~A. The only contributions from the model involved in organizing the data in panel~A into the master curve in panel C are the scaling-analysis estimates of $\phibar$  in Eqn.~\eqref{e:phibar} and  $\Wiebar$ in Table~\ref{tab:pipkin_summary}. The remaining values are from the experiments themselves. That data from systems differing so widely in chemistry, molecular weight, concentration and geometry should organize onto a single line -- using normalization factors that are obtained from a model strongly suggests that the C2D2 model has indeed largely captured the common mechanisms that underlie the different systems.

This reframes the original questions that prompted the current study. \citet{Gaillard2024-xu} found that a single solution, measured on devices of different plate radii, returned systematically different apparent relaxation times, and concluded that one should ``beware of CaBER.'' On the master plot, that plate-radius sweep is simply a traverse along the \(\Deve\) axis: the device geometry that appeared to make \(\lame\) ill-defined is nothing other than the input Deborah number, and varying it merely moves a sample along its Pipkin curve. The geometry dependence is thereby turned from a defect of the measurement into a controlled axis --- one along which the model's predictions and the experiments can be compared point by point. More tellingly, the same trend organizes data never suspected of such behaviour: the ``gold-standard'' Boger-fluid measurements of \citet{annamckinley} and \citet{clasenetal}, and the recent measurements of \citet{Calabrese2025-re}, whose authors reported no discernible geometry dependence at all, fall on the very line traced by Gaillard's geometry sweep, so that datasets which appeared to disagree prove mutually consistent once read through the framework.

There are three groups of outliers that do not neatly join the other points on the master curves. These are labelled in Fig.~\ref{fig:exptl} C as groups \(\mathrm{a}\), \(\mathrm{b}\) and \(\mathrm{c}\). Groups a and b  both lie well below the predicted master curve. This is because the model over-predicts \(\Wie\) relative to the data \citep{SI} so the renormalized values \( \Wie/\Wiebar \) fall below the plateau value of unity in panel C. The two groups sit in opposite concentration regimes and most likely have different origins for the deviations from the predicted master-curve. Group~\(\mathrm{a}\) lies in the moderately dilute regime, with points around and beyond the overlap concentration. The model predicts that the coil--stretch hysteresis window size is maximal at \( \phi = 1 \) and then begins to decrease as the system enters the semidilute regime. The hysteresis width, measured by the ratio \(\Wics/\Wisc\), is proportional to the ratio of the fully stretched to the coiled chain drag \citep{Prabhakar2016, Prabhakar2017-ri}. Beyond overlap the stretched-state drag changes little while the screening of intramolecular hydrodynamic interactions raises the coiled-state drag, so the ratio falls. While $\Wics$ remains at 1/2, $\Wisc$ increases. Therefore, the C2D2 model predicts that the plateau \(\Wiebar=(4/3)\Wisc\) should rebound upward along with $\Wisc$. The Boger fluid sample points in group-\(\mathrm{a}\) in panel C show no such rebound; their \(\Wie\) stays depressed. This could point to physics near the overlap concentration that the  blob arguments underlying the C2D2 model do not yet capture. Group~\(\mathrm{b}\), by contrast, lies in the highly dilute regime, where the elastic point is intrinsically hard to locate from the measured \(R(t)\) because the elastic signal is weak. It is possible therefore that the outliers in box b are due to limitations of the measurements rather than of the model. Understanding the third labelled outlier, group~\(\mathrm{c}\), requires an examination of the corner-shift values \(\Kstar\) for all the experimental systems.

Figure~\ref{fig:exptl_kstar} considers the corner shift values \( \Kstar \) across the seven experimental systems. The values span some two and a half decades, from \(\Kstar\approx0.003\) for the \citet{Hu2025-rt} PEO--PEG-8M solution to \(\Kstar\approx1.25\) for the \citet{Gaillard2024-xu} PEO solution; the remainder --- the PS Boger fluids and the \citeauthor{Hu2025-rt} PIB and PEO-1M solutions --- gather into a broad central band of roughly \(0.01\)--\(0.1\). If the experimental points had been tightly clustered along a single ray, it would have indicated that the blob-theoretic model and the scaling analysis accurately capture the underlying physics, and that the only remaining freedom is the set of order-unity prefactors that the blob-scaling arguments leave undetermined.  This would be expected only if all the experimental systems belonged to a single class sharing the same governing physics. The systems compared here share at least one such ingredient: the conformation-dependent hydrodynamic drag that the C2D2 model encodes, and that alone is enough to gather most of them together. The width around the mean value of $\Kstar = 0.07$ of the shaded band in Fig.~\ref{fig:exptl_kstar} is nevertheless broad, and two systems fall well outside that band. This could be due to two distinct causes.

The first is the crudeness of the model input values that the comparison presently rests on. Placing a system on the master plot calls for several molecular and geometric quantities that the original experiments, built to report an apparent relaxation time, had no reason to determine precisely. The intrinsic relaxation time \(\lamI\) is in several cases inferred indirectly --- for the \citet{Gaillard2024-xu} solution, for instance, from the solution viscosity of a fairly dilute sample rather than from an independent spectral measurement. The reduced concentration \(\phi=c/\cstar\) depends on an estimate of \(\cstar\), for which no single operational definition serves every system, so that different measures have had to be adopted across the datasets. We further treat each sample as monodisperse and assign it one \(\NK\) from its nominal molecular weight. Most consequentially, the onset radius \(\Rve\) that sets \(\Deve\) is seldom reported and has had to be read approximately from published radius--time curves; a direct record of strain rate against time, in which the minimum of the elastic phase can be read cleanly, would sharpen both \(\Rve\) and \(\Wie\). 

The second cause lies in the model. The C2D2 drag model captures the essential fact that a chain's average friction grows as it stretches, but the law has so far been tested only qualitatively, against multichain Brownian-dynamics simulations of polymer solutions in extensional flow \citep{Sasmal2016}. Its limitations are already visible in the group-\(\mathrm{a}\) Boger fluids near the overlap concentration, whose plateaus do not rebound as the model predicts, a point we return to below. Beyond such refinements the model omits physics that several of these systems possess. It contains no excluded-volume interactions, yet most of the solutions here are in better-than-theta solvents, and how sensitive the predicted plateau and corner are to solvent quality is not known. The aqueous PEO systems carry a further complication: intra- and intermolecular associations arising from hydrogen bonding are known to shape the solution behaviour of PEO in water \citep{Ebagninin2009}, and are absent from the model altogether. It is telling that \citet{Gaillard2024-xu} found their own PEO solutions to age, their apparent relaxation time and zero-shear viscosity falling by some thirty percent over several months. The extent to which such interactions will modify model predictions is unknown. Each of these missing ingredients could contribute to a departure of \(\Kstar\) from the value the drag physics alone sets.

Of the two systems that fall outside the central band, the PEO solution of \citet{Gaillard2024-xu} at \(\Kstar\approx1.25\) is plausibly brought there by the two effects just described: an indirectly inferred \(\lamI\) compounded by the associative peculiarities of aqueous PEO. The low-side outlier is not so easily accommodated. The 8M-PEO solution of \citet{Hu2025-rt} is also the outlier group~\(\mathrm{c}\) in Fig.~\ref{fig:exptl} C: its \(\Kstar\approx0.003\) lies below the lower edge of the band, more than an order of magnitude beneath its centre.  The lower-molecular-weight 1M-PEO solution in that study shows no such anomaly and collapses with the other systems. While the reasons behind the deviation in the 8M-PEO solution are not clear, the evidence points to chain scission. The strain-rate histories reconstructed in the Supplementary Information \citep{SI} from the raw \(R(t)\) data of \citeauthor{Hu2025-rt} show the neck sustaining peak Weissenberg numbers of order \(10^2\) through the approach to the elastic onset, reached in both the dripping and the dripping-onto-substrate runs; since \citeauthor{Hu2025-rt} deliberately held the injection flow slow to avoid pre-stretching the chains, the exposure to high \(\Wi\) during the pre-elastic thinning itself is common to the two very different bridge-formation protocols used to obtain the 8M-PEO solution data.   It has been suggested that chains can undergo scission in strong extensional flows \citep{Odell1990-so, Buchholz2004-mh}. \citet{gupta} observed steady-state \emph{extension thinning} in filament-stretching rheometry, in which beyond \(\Wi\approx10\) the extensional viscosity of PS Boger fluids \emph{decreases} with increasing strain rate; this behaviour has been attributed to chain scission by \citet{larsonreview}.  \citet{Joseph2025-ls} have recently reported further evidence for scission in capillary-breakup extensional rheometry specifically, particularly in Dripping-on-Substrate devices.

\section{Discussion}
\label{sec:discussion}

A master-curve collapse as that in Fig.~\ref{fig:exptl} C may cause one to wonder: might it not be an artifact of the normalization itself? If the scaling factors carried enough adjustable information, perhaps \emph{any} scattered cloud of points could be marshalled onto a line, so that the apparent collapse would say nothing about the physics. This cannot be so, for two reasons. First, the normalization factors  \(\Wiebar\), \(\Devebarpred\) are \emph{predicted}  and the master shape follows from the C2D2 scaling analysis evaluated at \(\phi\), \(\NK\) and \(\hsK\) fixed independently of the master plot. The only per-system freedom left  is the single horizontal shift applied to a whole dataset corresponding to a single molecular weight in a given polymer-solvent combination. This shift cannot repair a master curve of the wrong shape. Second, and decisively, a model with the wrong microphysics simply produces the wrong factors. The constant-drag FENE-P model assigns \(\Wiebar=2/3\) to every system and provides no concentration-dependent corner; it therefore cannot account for the many experimental points that lie \emph{below} \(2/3\), and no application of its scaling factors would bring them onto a common line. Only a model whose physics yields the correct plateau and corner -- here the conformation- and concentration-dependent drag and the coil--stretch hysteresis it encodes -- can rationalize the raw Pipkin data of Fig.~\ref{fig:exptl} A and collapse them onto the master line of Fig.~\ref{fig:exptl} C.

That C2D2 does so is, given the crudeness of our inputs, a strong result. The collapse spans systems differing by decades in molecular weight and concentration, by solvent quality, and by bridge geometry, yet rests on order-of-magnitude estimates of \(\NK\) and the relaxation times and a single hand-tuned \(\hsK\) per system. It is in this sense that CBR experiments become a stringent  benchmark for microstructural models: the quality of the collapse measures directly the fidelity of the microstructure model, and would do so for any other constitutive class -- semidilute entangled solutions, polyelectrolytes, associating polymers, \emph{etc.} -- tested the same way. 

\citet{Hu2025-rt} analyse the same body of geometry- and size-dependent apparent relaxation times in the experiments and also demonstrate data collapse onto master curves. The difference between the two is in how the data is treated and interpreted. For each experimental system consisting of a solvent and polymer of given molecular weight, \citeauthor{Hu2025-rt} take the constant-drag FENE-P model as ground truth, use its relaxation time  and finite-extensibility parameter (which is, in principle, proportional to \(\NK\)) as tuneable constants to match that model's predictions to the experimentally-observed variation of the apparent relaxation time \(\lame\) with the device-geometry-dependent initial filament radius, \( \Ro \).  The \(\Ro\)-independent \(\lambda_\mathrm{fit}\)  and \(b_\mathrm{fit}\)  thus extracted are claimed to be intrinsic molecular properties of the polymer solution. Then, \(\lame/\lambda_\mathrm{fit}\) for different experimental systems is plotted against an elastocapillary number, \(\mathrm{Ec}_e=\gamma\lame/(\Ro \etap)\), where \(\etap = \eta_0-\etas\) is the polymer contribution to the zero-strain-rate shear viscosity of the solution. To within the Entov--Hinch 2/3 factor, their ordinate is the reciprocal of \(\Wie\);  their master-plots  are our Pipkin diagrams vertically inverted, with our low-\(\Deo\) plateau mapping to a plateau of theirs and our post-corner rise comparable to the power-law branch they observe in their master curves. 

However, where the  near-equilibrium (SAOS) relaxation time \(\lamI\) data are available, the recovered \(\lambda_\mathrm{fit}\) departs from  it in \emph{both} directions (below it for PS--DOP, above it for PIB--PB). That leaves the central question: what does \( \lambda_\mathrm{fit} \) physically mean? Without a clear answer, one cannot say how this quantity ``characterizes'' a polymer solution in the unambiguous sense that \(\lamI\) does. Ultimately, the real power of \(\lamI\) in characterizing polymer solutions is that we understand well --- through well-tested microstructural models --- how it is connected to the more fundamental molecular parameters, the molecular weight and the polymer concentration. The source of the ambiguity in the meaning of  \( \lambda_\mathrm{fit} \) lies in force-fitting a model without adequate microstructural physics to the experimental data.

The C2D2 model used in the present study does incorporate  additional microstructural physics. The normalization factors that are required to organize the data into the master curve in Fig.~\ref{fig:exptl} C are obtained systematically from a scaling analysis.  The extra shift factor \( \Kstar \) that is required over and above the scaling-derived normalization factors to collapse the data  can be systematically probed and attributed, either to lack of experimental detail or to the physics the microstructural model still lacks. If one assumes that the empirical shift \(\Kstar\) is solely due to the assumption of \(O(1)\) prefactors in the blob model or the scaling analysis,  one would like to trace it to its source in the model, so that the corrected normalization factors \( \Wiebar \) and \(\Devebarpred\) would carry it automatically. Where the correction should enter in the scaling analysis of Appendix~\ref{app:scaling-derivation} is however currently unclear: the corner and the plateau draw on overlapping molecular quantities, and the two Pipkin-curve families weight them differently, so that no single plateau-preserving factor reproduces a common shift across all systems.  That in itself may point to a real gap in the C2D2 description.

The model-versus-experiment comparison here also points to another feature that is poorly understood: the anomalously small values of the HI parameter  \(\hsK\) (also noted previously by  Prabhakar and co-workers \citep{Prabhakar2016, Prabhakar2017-ri, Wang2025-sf}). This parameter is proportional to the hydrodynamic radius of a Kuhn segment relative to its length, and it is not clear why the values required to describe experimental data are much smaller than the \( O(1) \) values expected. For solutions of long, flexible chains, this parameter is difficult to obtain from fits of the Zimm model's predictions to SAOS spectra. In such fits, one typically uses the most convenient value for the HI parameter of a coarse-grained spring to describe the experimentally observed spectra with the least number of modes. Such fits yield values  around 0.2 for the $h^\ast$ of the coarse-grained segments. The origin of this near-universality is well understood \citep{Kroger2000}. Flow-induced stretching or inter-chain screening makes  rheological behaviour  more sensitive to local parameters such as $\hsK$. Thus, CBR experiments can be used to extract their values, and explore and better understand their systematic dependence on polymer-solvent chemistry.

The analysis here has been confined to the viscocapillary regime, in which the pre-elastic balance is dominated by the solvent-viscous stress. When inertia is no longer negligible---the inertio-capillary regime relevant to low-viscosity solvents---the pre-elastic stress balance changes, and with it the power-law exponents that set the affine branch, the corner, and the post-corner rise. We therefore expect the numerical values of the scaling exponents to shift in that regime. The selection mechanism itself, however, is expected to remain unchanged: the elastic point is still set where the snapback segment of the transient trajectory meets the steady-state constitutive curve, on either its coil-stretch-transition or its finite-extensibility branch. Just as in the viscocapillary case here, we can expect Pipkin curves to have a plateau and a post-corner rise. How \(\Wie\) depends on \(\Deo\) or \(\Deve\) there is dictated by the power-law growth of the polymer viscosity with \( \Wi \) in the pre-elastic phase, which in turn will depend on the pre-elastic inertio-capillary balance. The scaling analysis framework developed here can be used to derive the asymptotic power-law exponent, as well as estimates for $\phibar$, $\Debaro$ and $\Devebar$ for the inertiocapillary case as well.

The master-curve collapse in Fig.~\ref{fig:exptl} C is encouraging, but the discussion above also suggests how CBR experiments should be designed to provide datasets that serve as impartial benchmarks for microstructural models, in the same manner that FiSER data are today used to benchmark models for entangled polymer solutions.  In light of the framework set out here, a purpose-built study must pin down, for each sample, four quantities. The intrinsic relaxation time \(\lamI\) should be measured independently, by small-amplitude oscillatory shear, rather than inferred from the thinning itself; the concentration should be referred to an overlap \(\cstar\) defined consistently through the Martin coefficient, so that \(\phi\) is comparable across chemistries; the sample should be well characterized and near-monodisperse, so that a single \(\NK\) is meaningful; and, most consequentially, the neck evolution should be resolved finely enough --- ideally as a strain rate against time, not merely a radius against time --- that the elastic-onset radius \(\Rve\) and the strain-rate at the elastic point, \(\edote\), can be read cleanly.

The choice of system and apparatus matters just as much. The earliest CBR devices used rapid separation of end-plates to create liquid bridges. To suppress the mechanical instability triggered by the rapid initial motion of the end plates in this step-strain mode necessitated the use of very viscous solvents \citep{McKinley2005-fl}. Techniques such as the slow-retraction method  of \citet{CDeano2010} or the dripping-on-substrate method  allow one to handle much lower viscosity solvents. This opens up the systematic and accurate exploration of solutions of polymers in molecularly simple solvents, such as the PS-in-DOP  system studied by \citet{Calabrese2025-re}. Boger fluids may have complex polymer-solvent interactions that are not fully understood \citep{Lee1999-pj, Solomon1996b}. This system also avoids the associative interactions suspected in PEO solutions in water-based solvents. With careful temperature and evaporation control \citep{Robertson_Calabrese, Hu2025-rt}, one could also explore the influence of solvent quality.  Moreover, since the model spans the dilute and semidilute regime, there is no fundamental reason to confine such studies to ultra-dilute where estimations of the elastic point become difficult. One can operate at concentrations around critical overlap \(\phi\sim0.1\)--\(3\). This is also essential to understand the crossover to semidilute behaviour.  A programme of this kind would allow the effect of chain scission to be  systematically studied.   Since data on a single solution measured on different devices should, in principle, collapse onto one master curve, running the same sample through SR-CaBER, DoS, and droplet-dripping geometries --- which impose different bridge-formation and pre-elastic histories --- would test that device-independence directly. Large deviations, and systematic non-repeatability following sample resting, could be attributed to chain scission. 

 Once the mechanism proposed in the present study is experimentally verified, it paves the way for true microstructural characterization of polymer solutions. Instead of relying on ``apparent relaxation time'' estimates of doubtful value,  given high-quality \(R(t)\) data, known solvent properties, and a well-tested model such as the C2D2 framework, one may develop systematic protocols to infer underlying microscopic parameters -- including molecular weight, concentration, intrinsic relaxation time, and local parameters such as the Kuhn length and $\hsK$ -- in a manner consistent with the full thinning dynamics rather than through a single-point measurement such as $\Wie$.  A thinning curve \(R(t)\) --- and, better still, a family of them measured across geometries and concentrations --- carries far more information than the single number \(\Wie\): the pre-elastic viscocapillary stage, the elastic onset, and the post-corner rise each constrain a different combination of the molecular parameters. Treating the C2D2 forward model as the likelihood and the molecular parameters as unknowns with physically-motivated priors, a Bayesian inversion would return full posterior distributions rather than point estimates, with   uncertainties and with the parameter degeneracies  made explicit rather than buried in a single best-fit number.  The open-source code made available with this work is intended to make such developments possible \citep{c2d2jl,c2d2report}.

The approach developed here could further be extended to understand the observations of \citet{Aisling2024-ie} on the effect of moving the end plates continously at a steady velocity.  This introduces the Weber number corresponding to the end-plate motion as an extra parameter, appearing as a boundary condition. The analysis underlying normal-stress extraction in FiSER suggests that the stress balance can be adapted to model the macroscopic flow at the neck of such a device. This flow  then becomes the \emph{next} ``simplest complex flow'' coupling macroscopic dynamics to microscopic evolution,  one further step along a  spectrum of complex flows that runs from CBR with static end-plates, through devices with moving endplates such as that of \citeauthor{Aisling2024-ie} and FiSER, and toward more complex 2D and 3D fully inhomogeneous benchmark flows for which constitutive models are ultimately needed.

\section{Conclusions}
\label{sec:concl}

The apparent relaxation time \(\lame\) that capillary-breakup rheometry is usually asked to deliver is not an intrinsic material property: it varies with concentration, solvent quality, bridge geometry and preparation protocol. We have argued that this variability is not a defect of the experiment but a failure of the Entov-Hinch prediction \(\Wie=2/3\), on which the extraction of \(\lame\) rests. That prediction was based on a constitutive model that assumed a constant relaxation spectrum. Although many find the constant-drag FENE-P model with a fixed relaxation time convenient to work with, the large body of work on conformation-dependent intra- and intermolecular hydrodynamic interactions has shown that that is not the correct microstrictural picture, including work on capillary thinning \citep{Ottinger87,magda,Ottinger89,Kishbaugh1990,Ottinger92,prabhakar_gaussian,prabhakar,prabhakarSFG,Prabhakar2016,Prabhakar2017-ri,Wang2025-sf}. Therefore, thinking about CBR observations in terms of \(\lame\) is perhaps best avoided.

The remedy is to change what one reads from the experiment. CBR is best viewed as stress-controlled extensional rheometry: the thinning neck imposes a capillary stress and the strain rate at the neck is selected, not prescribed. Its model-agnostic output is that selected rate, expressed through the elastic Weissenberg number \(\Wie\) --- the counterpart of the normal-stress and extensional-viscosity outputs of a filament-stretching rheometer --- and, like them, it acquires meaning only alongside an independently measured intrinsic relaxation time \(\lamI\).

Read this way, the geometry and protocol dependence that made \(\lame\) seem ill-defined becomes the signal to be modelled. Plotting \(\Wie\) against the geometry-controlled Deborah number \(\Deo\) yields elastocapillary Pipkin diagrams whose curves --- a low-\(\Deo\) plateau, a corner, and a finite-extensibility rise --- are organized by a scaling analysis in the polymer-viscosity--Weissenberg-number phase plane. The conformation- and concentration-dependent drag of the C2D2 model, acting through coil--stretch hysteresis, is what lowers the plateau below the Entov--Hinch value of \(2/3\), as the experiments require; and a renormalized Deborah number \(\Deve\) absorbs the unmeasured initial prestretch into a single observable. With these model-predicted factors, CBR data spanning decades in molecular weight and concentration, several chemistries, and a range of devices --- including the apparently troublesome geometry sweep of \citet{Gaillard2024-xu} --- collapse onto a single master curve.

That such disparate data organize onto one line, using normalizers a model supplies, is what makes CBR a stringent test of microstructural physics. Because its dynamics localize to the necking plane, capillary thinning is perhaps the \emph{simplest complex flow} a polymer solution can be subjected to, coupling the macroscopic balance to the evolving microstructure without spatial inhomogeneity. A constitutive model that cannot describe it is unlikely to fare better in the inhomogeneous flows for which it is ultimately intended: to win, a model must first thin.

The collapse is not perfect. The moderately dilute points near the overlap concentration, whose plateaus fail to rebound as predicted, and the order-unity offset absorbed into the empirical corner shift, both mark physics the blob-based C2D2 model does not yet contain. Rather than weaknesses of the method, these are the openings it creates: CBR, read as a benchmark, turns each such deviation into a quantitative target for refining microstructural models of dilute and semidilute solutions. The open-source implementation and technical note released with this work are intended to support that programme \citep{c2d2jl, c2d2report}.

\appendix

\section{Scaling analysis of the stress-balance equation}
\label{app:scaling-derivation}

This appendix supplies the scaling arguments behind the plateau and corner values quoted in the main text (Table~\ref{tab:pipkin_summary}). We revisit and extend the arguments of \citet{entovhinch} to explain how the geometry-controlled Deborah number \(\Deo\) sets the selected elastic Weissenberg number \(\Wie\). Although the discussion is a scaling analysis, we retain numerical prefactors where they arise naturally and are required to ensure continuity across scaling regimes.

\subsection{The pre-elastic phase}

At early times the polymer normal stress is negligible, so capillarity is balanced by the viscous solvent stress. With \(\NIc=(2X-1)\Deo/R\) and \(\NIp=\UI\phi\tNIp\) the dimensionless capillary and polymer normal stresses (each scaled by \(\etas/\lamI\)), and the solvent-viscous resistance \(\NIv=3\Wi\), the pre-elastic balance gives
\begin{gather}
\Wi(t)  \simeq \frac{\Hv\,\Deo}{R(t)},
\label{e:Wit-visc}
\end{gather}
where \(\Hv=0.142\) is the geometric coefficient introduced in Sec.~\ref{sec:mfsbe}. At \(t=0\), \(R(0)=1\) and \(\NIp=0\), recovering the initial value \(\Wio=\Hv\Deo\) of Eq.~\eqref{e:Wio}.

\paragraph{Initially-weak and initially-strong stretching.} Whether the chains start weakly or strongly stretched is set by \(\Wio\) relative to the coil--stretch threshold \(\Wics\): \emph{initially-weak stretching} when \(\Wio<\Wics\) (equivalently \(\Deo<\Wics/\Hv\)) and \emph{initially-strong stretching} when \(\Wio>\Wics\). 

The Weissenberg number at which \emph{affine stretching begins} is therefore
\begin{gather}
\Wi_\mathrm{in} = \max(\Wics,\Wio),
\label{e:wii}
\end{gather}
namely \(\Wics\) for initially-weak and \(\Wio\) for initially-strong stretching. At fixed \(\phi\), every initially-weak stretching trajectory creeps along the zero-extension-rate plateau \(\teetapo\) to the same point \(\Wi=\Wics\) before stretching, tracing a common path -- the orange trajectory in Fig.~\ref{fig:scaling_analysis}B\,(i) and the green in B\,(ii) -- independent of \(\Deo\). Only once \(\Deo>\Wics/\Hv\) do the initially-strong stretching trajectories (red in B\,(i), purple in B\,(ii)) shift rightward with increasing \(\Deo\).

\paragraph{Affine stretching.} Along any single trajectory, once \(\Wi\gtrsim \Wi_\mathrm{in}\), chains stretch affinely. Finite extensibility is still unimportant, so that \(f\simeq 1\). From Eqs.~\eqref{e:kramers} and \eqref{e:teetapdef}, we expect the scaling \(\teetap \sim M/\Wi\), which we write with consistent prefactors as
\begin{gather}
\teetap \simeq \teetapo\,\Wics\,\frac{M}{\Wi},
\label{e:teetap-affine}
\end{gather}
with the understanding that \(\Wi \ge \Wi_\mathrm{in}\) and \(M \ge 1\). The affine conformation result is purely kinematic: the conformation equations give \(d\ln\Mzz/dt\simeq-4\,d\ln R/dt\) and \(d\ln\Mrr/dt\simeq 2\,d\ln R/dt\), so that
\begin{gather}
\Mzz \simeq \tfrac{1}{3}R^{-4},
\qquad
\Mrr \simeq \tfrac{1}{3}R^2,
\qquad
M \simeq \Mzz .
\label{e:Mzzviscregime}
\end{gather}
Combining \(\Mzz\sim R^{-4}\) with the pre-elastic relation \(\Wi\sim 1/R\) [Eq.~\eqref{e:Wit-visc}] gives the affine-stretching branch
\begin{subequations}
\label{e:viscousaffine}
\begin{align}
M &\simeq \left(\frac{\Wi}{\Wi_\mathrm{in}}\right)^{4}, \label{e:viscousaffine-M}\\[2mm]
\teetap &\simeq \teetapo\,g_\mathrm{in}\left(\frac{\Wi}{\Wi_\mathrm{in}}\right)^{3}, \label{e:viscousaffine-eta}
\end{align}
\end{subequations}
where
\begin{gather}
g_\mathrm{in} = \min\!\left(1,\frac{\Wics}{\Wio}\right).
\label{e:gi}
\end{gather}
Thus both initially-weak and initially-strong stretching trajectories exhibit the same scaling forms \(M\sim \Wi^4\) and \(\teetap\sim \Wi^3\) -- the slope-\(3\) \emph{affine stretching} segments in the lower panels of Fig.~\ref{fig:scaling_analysis} -- differing only in the point at which the affine branch is entered.

\subsection{The elastic onset}

The elastic onset is the instant \(t=t_{*}\) at which the growing polymer stress first matches the solvent-viscous resistance, \(\NIp \sim \NIv\). Two numbers label it: the onset Weissenberg number \(\Wive\), and the onset stretch \(M_*\) that affine stretching has reached by that instant. We mark onset quantities with an asterisk: in earlier work the onset has commonly carried the subscript~\(1\), but we reserve that subscript for the longest relaxation time \(\lamI\), and so write \(\Wive\), \(M_*\), and \(t_*\). The affine branch~\eqref{e:viscousaffine-M} ties the two together,
\begin{gather}
M_* = \left(\frac{\Wive}{\Wi_\mathrm{in}}\right)^{4}.
\label{e:Mstar}
\end{gather}
The onset occurs when the polymer and solvent-viscous contributions become comparable. With \(\NIv=3\Wi\), \(\NIp=\UI\phi\,\tNIp\), and \(\teetap=\tNIp/\Wi\), the onset viscosity is thus  \(\Deo\)-independent,
\begin{gather}
\teetapve \simeq \frac{3}{\UI\,\phi}.
\label{e:teetapve}
\end{gather}
Substituting Eq.~\eqref{e:teetapve} into the affine branch~\eqref{e:viscousaffine-eta} fixes the onset Weissenberg number,
\begin{gather}
\Wive
\simeq
\Wi_\mathrm{in}\left(\frac{3}{\teetapo\,g_\mathrm{in}\,\UI\,\phi}\right)^{1/3}.
\label{e:wive}
\end{gather}

\paragraph{Weak initial stretching.}
For \(\Wio<\Wics\) the chains reach the coil--stretch threshold before they stretch appreciably, so \(\Wi_\mathrm{in}=\Wics\) and \(g_\mathrm{in}=1\). The onset Weissenberg number is then independent of \(\Deo\) and falls with concentration,
\begin{gather}
\Wive
\simeq
\Wics\left(\frac{3}{\teetapo\,\UI\,\phi}\right)^{1/3}
\propto \phi^{-1/3}.
\label{e:wive-weak}
\end{gather}

\paragraph{Strong initial stretching.}
For \(\Wio=\Hv\Deo>\Wics\) affine stretching begins at \(t=0\), so \(\Wi_\mathrm{in}=\Wio=\Hv\Deo\) and \(g_\mathrm{in}=\Wics/\Wio\). The onset then shifts rightward with the geometry: Eq.~\eqref{e:wive} gives \(\Wive\sim\Wi_\mathrm{in}^{4/3}\), i.e.
\begin{gather}
\Wive \sim (\Hv\Deo)^{4/3}\,(\teetapo\,\UI\,\phi\,\Wics)^{-1/3}
\propto \Deo^{4/3}\,\phi^{-1/3}.
\label{e:wive-strong}
\end{gather}
The same \(\phi^{-1/3}\) governs both branches, so the weak and strong forms match at the crossover \(\Deo=\Wics/\Hv\); the strong branch differs only in acquiring the \(\Deo^{4/3}\) geometric growth. From Eq.~\eqref{e:Mstar}, the onset stretch then climbs as \(M_*\sim(\Hv\Deo)^{4/3}\) along this branch.

\subsection{Snapback follows \(\teetap\sim\Wi^{-1}\)}

Once polymer stresses dominate, the dynamics switch from the pre-elastic balance to a quasistatic elastocapillary balance. The self-selected \(\Wi\) on that branch is always much smaller than \(\Wive\), so the transient trajectory undergoes a sharp drop in \(\Wi\) while \(M\) changes only weakly. Since the polymer stress remains approximately fixed during this short interval, the transient extensional viscosity scales inversely with \(\Wi\), and the snapback segment obeys
\begin{gather}
\teetap \simeq \teetapve\left(\frac{\Wive}{\Wi}\right).
\label{e:eetap-snapb}
\end{gather}
This relation is purely a consequence of fixed polymer stress, and traces the slope-\(-1\) \emph{snapback} segments in the lower panels of Fig.~\ref{fig:scaling_analysis}.

\subsection{Elastic points}

The elastic point -- the turning point where the transient \(\Wi\) reaches its minimum \(\Wie\) -- is selected where the snapback segment meets the steady-state curve. Two cases follow, according to which branch of the steady curve it meets.

\paragraph{SCT-limited elastic points.}
If, at the elastic onset, \(M\ll \NK\) so that finite extensibility remains unimportant, then with \(M\simeq M_{zz}\) the elastocapillary balance \(\NIp\simeq \NIc\) implies
\begin{gather}
\UI\,\phi\,\teetapo\,\Wics\,M \simeq \frac{\Hv\,\Deo}{R},
\label{eq:balance-el}
\end{gather}
whose logarithmic derivative gives \(d\ln M/d\varepsilon \simeq 1/2\), with \(\dot\varepsilon=-2\,d\ln R/dt\). On the other hand, at large stretch the conformation dynamics in strain form give \(d\ln M/d\varepsilon \simeq 2-1/(\nu\Wi)\). Equating the two yields the self-selected Weissenberg number in the affinely extensible elastic regime, \(\Wi \simeq (2/3)/\nu\), while the steady solution at the same \(M\) is \(\Wi_{\rm ss}\simeq(1/2)/\nu\). Hence the transient follows the retrograde steady branch with the fixed offset \(\Wi\simeq(4/3)\Wi_{\rm ss}\). If the snapback segment reaches the steady-state curve before finite-extensibility effects become important -- as for the green and blue trajectories of Fig.~\ref{fig:scaling_analysis}B\,(ii) -- the turning point is SCT-limited and occurs where \(\Wi_{\rm ss}=\Wisc\), giving
\begin{gather}
\Wie = \frac{4}{3}\Wisc.
\label{e:wie-aee}
\end{gather}
In the FENE-P model, \(\Wisc=\Wics=1/2\) and this reduces to the Entov--Hinch result \(\Wie=2/3\).

\paragraph{FE-limited elastic points.}
We define the onset of finite-extensibility-dominated behaviour by \(M \simeq \FE \NK\), where \(\FE\) is an empirical coefficient that we estimate to be \(0.22\). By this stage the drag ratio is close to that of a highly stretched chain, \(\nu_\fs=\Wics/\Wisc\) \citep{Prabhakar2016}. Evaluating Eq.~\eqref{e:teetap-affine} at \(M\simeq\FE \NK\) and \(\Wi\simeq(2/3)/\nu_\fs\) gives the extensional viscosity above which finite-extensibility effects matter,
\begin{gather}
\teetapfe
\equiv
\frac{\teetapo\,\Wics\,\FE \NK}{(2/3)/\nu_\fs}
=
\frac{3}{2}\,\teetapo\,\Wics\,\FE \NK\,\nu_\fs.
\label{e:teetapfe}
\end{gather}
If the snapback segment instead lands on this finite-extensibility-dominated branch -- as for the orange and red trajectories of panel B\,(i) and the purple trajectory of panel B\,(ii) -- the elastic point is FE-limited. Equating the snapback relation \eqref{e:eetap-snapb} with \(\teetapfe\) gives \(\Wie=(\teetapve/\teetapfe)\,\Wive\). Using \(\teetapve\Wive=\teetapo\Wics M_{*}\) with the affine stretch at onset \(M_{*}=(\Wive/\Wi_\mathrm{in})^{4}\), the elastic point is set by the onset stretch, \(\Wie\propto M_{*}\). For the initially-weak stretching plateau, \(M_{*}\propto(\phi^{-1/3})^{4}=\phi^{-4/3}\). Substituting Eqs.~\eqref{e:teetapve}, \eqref{e:wive}, \eqref{e:teetapfe} then yields
\begin{gather}
\Wie
=
\frac{2}{3}\,\frac{\Wisc}{\Wics}
\left[
\frac{3^{4/3}}{\FE \NK}
\,
\frac{\Wi_\mathrm{in}/\Wics}{(\teetapo \UI\phi)^{4/3} g_\mathrm{in}^{1/3}}
\right].
\label{e:wie-fee}
\end{gather}
This is the FE-limited value of the elastic Weissenberg number.

\subsection{The boundary concentration \(\phibar\) and the corner \(\Debaro\)}

For the initially-weak stretching trajectories, \(\Wi_\mathrm{in}=\Wics\) and \(g_\mathrm{in}=1\). These trajectories therefore all share the same FE-limited plateau value,
\begin{gather}
\Wiebar
=
\frac{2}{3}\,\frac{\Wisc}{\Wics}
\left[
\frac{3^{4/3}}{\FE \NK\,(\teetapo \UI\phi)^{4/3}}
\right],
\label{e:wiebar-felim}
\end{gather}
and the common corner value, fixed by the initially-weak--to--initially-strong crossover \(\Wio=\Hv\Deo=\Wics\),
\begin{gather}
\Debaro = \frac{\Wics}{\Hv}.
\label{e:debar-felim}
\end{gather}
These correspond to the highly dilute, FE-limited Pipkin curves -- the upper, orange-to-red curve of Fig.~\ref{fig:scaling_analysis}A\,(i), whose plateau \(\Wiebar\) sits above \(2/3\) and whose corner \(\Debaro=\Wics/\Hv\) is independent of \(\phi\). The constant plateau reflects the common initially-weak stretching path traced by the orange trajectory of panel~B\,(i).

The boundary concentration \(\phibar\) between the highly dilute and moderately dilute families is obtained by equating the FE-limited plateau \eqref{e:wiebar-felim} to the SCT-limited value \eqref{e:wie-aee}. This is equivalent to setting the square bracket in Eq.~\eqref{e:wiebar-felim} to unity, which gives Eq.~\eqref{e:phibar}.

For the moderately dilute, SCT-limited family, the corner \(\Debaro\) is determined by asking when the finite-extensibility correction factor in Eq.~\eqref{e:wie-fee} becomes \(O(1)\) along the initially-strong stretching branch. Substituting \(\Wi_\mathrm{in}=\Wio=\Hv\Deo\) and \(g_\mathrm{in}=\Wics/\Wio\) into Eq.~\eqref{e:wie-fee}, the bracket becomes
\begin{gather}
\frac{3^{4/3}}{\FE \NK}
\left(
\frac{\Hv\Deo}{\Wics\,\teetapo \UI\phi}
\right)^{4/3}.
\end{gather}
Setting this equal to unity at \(\Deo=\Debaro\) yields
\begin{gather}
\Debaro
=
\frac{\Wics}{\Hv}\,
\frac{\teetapo \UI\phi\,(\FE \NK)^{3/4}}{3}
=
\frac{\Wics}{\Hv}\,\frac{\phi}{\phibar}.
\label{e:debar-trans}
\end{gather}
Hence the corner shifts linearly with concentration in the SCT-limited family -- the lower, blue/green-to-purple curve of Fig.~\ref{fig:scaling_analysis}A\,(i), whose corner moves to larger \(\Deo\) as \(\phi\) increases, driven by the rightward shift of the initially-strong stretching (purple) trajectories in panel~B\,(ii).

\subsection{The post-corner rise and the operating-window boundaries}

Once \(\Deo>\Debaro\), the elastic point is FE-limited in both Pipkin families. Inserting the initially-strong stretching values \(\Wi_\mathrm{in}=\Hv\Deo\) and \(g_\mathrm{in}=\Wics/(\Hv\Deo)\) into Eq.~\eqref{e:wie-fee} shows that
\begin{gather}
\Wie \sim \Deo^{4/3},
\label{e:postcorner}
\end{gather}
the slope-\(4/3\) post-corner rise marked in Fig.~\ref{fig:scaling_analysis}A\,(i).

As \(\phi\) and \(\Deo\) vary, the elastic onset traces a locus in the \((\Wi,M)\) plane. Read against the finite-extensibility ceiling \(M=\NK\), this locus decides whether an elastic regime exists at all: the two boundaries below are the points at which it meets the ceiling.

\paragraph{The maximum Deborah number \(\Deo^{\max}\).}
The onset stretch climbs along the initially-strong stretching branch, \(M_*\sim(\Hv\Deo)^{4/3}\), until it reaches the finite-extensibility ceiling. At \(M_*\simeq \NK\) the elastic onset coincides with the onset of finite extensibility, so no distinct elastic regime survives above it. Imposing \(M_*=\NK\) gives
\begin{gather}
\Deo^{\max} \sim \frac{\Wics}{\Hv}\,\teetapo\,\UI\,\phi\,\NK^{3/4}
\;=\; \FE^{-3/4}\,\Debaro,
\label{e:De0max}
\end{gather}
which grows linearly in \(\phi\), like the corner itself. In the moderately dilute family this is \(\Deo^{\max}\simeq\FE^{-3/4}\Debaro\approx 3\,\Debaro\): the post-corner window spans only about half a decade in \(\Deo\) before finite extensibility caps it. \(\Deo^{\max}\) closes the Pipkin diagram from the upper right, where the operating curve terminates.

\paragraph{The minimum concentration \(\phi_{\min}\).}
The complementary limit is the dilute floor. For \(\Deo\le\Wics/\Hv\) all trajectories share the common weakly-stretched path, so the onset stretch sits at its \(\Deo\)-independent plateau value \(M_*\simeq(\Wive/\Wics)^{4}\) with \(\Wive\) from Eq.~\eqref{e:wive-weak}. Imposing \(M_*=\NK\) gives the concentration below which the onset never reaches the finite-extensibility ceiling and no elastic regime exists,
\begin{gather}
\phi_{\min} \sim \frac{3}{\teetapo\,\UI\,\NK^{3/4}}
\;=\; \FE^{3/4}\,\phibar .
\label{e:phimin}
\end{gather}

\bibliography{bib_velbgeometry_v2}
\bibliographystyle{plainnat}

\end{document}

%% file: macros_velbgeometry.tex
\renewcommand{\vec}[1]{\mathbf{#1}}
\newcommand{\tr}{\mathrm{tr}}
\newcommand{\ut}{\mathbf{\updelta}}
\newcommand{\etas}{\eta_\mathrm{s}}

\newcommand{\etap}{\eta_\mathrm{p}}

\newcommand{\kBT}{k_{\textsc{b}} \,T}
\newcommand{\ckBT}{c\,k_{\textsc{b}} \,T}

\newcommand{\edote}{\dot{\varepsilon}_\mathrm{e}}

\newcommand{\Rb}{R_\mathrm{b}}
\newcommand{\Ro}{R_0}
\newcommand{\Rve}{R_{\ast}}   % prints R_* to match \Wive (Wi_*) and \Deve (De_*); renamed from \Rstar 2026-07-04
\newcommand{\Lb}{L_\mathrm{b}}
\newcommand{\Vb}{V_\mathrm{b}}

\newcommand{\Wi}{\mathrm{W\!i}}

\newcommand{\Wie}{\mathrm{W\!i}_{\,\mathrm{e}}}

\newcommand{\Wiebar}{\overline{\mathrm{W\!i}}_{\,\mathrm{e}}}
\newcommand{\Wive}{\mathrm{W\!i}_{\ast}}
\newcommand{\Wio}{\mathrm{W\!i}_{\,0}}
\newcommand{\Wics}{\Wi_\cs}
\newcommand{\Wisc}{\Wi_\sc}

\newcommand{\De}{\mathrm{De}}
\newcommand{\Deo}{\mathrm{De}_0}
\newcommand{\Debaro}{\overline{\De}_0}
\newcommand{\Oh}{\mathrm{Oh}}

\newcommand{\Bo}{\mathrm{Bo}}

\newcommand{\NK}{N_\textsc{k}}

\newcommand{\hsK}{h^\ast_\textsc{k}}

\newcommand{\NIp}{N_{1,\,\mathrm{p}}}
\newcommand{\NIv}{N_{1,\,\mathrm{v}}}

\newcommand{\NIc}{N_{1,\,\mathrm{c}}}

\newcommand{\tcap}{\lambda_\mathrm{v}}

\newcommand{\lame}{\lambda_\mathrm{e}}

\newcommand{\lamI}{\lambda_1}

\newcommand{\Hv}{H_\mathrm{v}}

\newcommand{\Xv}{X_\mathrm{v}}
\newcommand{\Xe}{X_\mathrm{e}}

\newcommand{\bM}{\vec{M}}

\newcommand{\bkappa}{\vec{\upkappa}}

\newcommand{\Ree}{E}
\newcommand{\Reo}{E_0}

\newcommand{\Mzz}{M_{zz}}
\newcommand{\Mrr}{M_{rr}}

\newcommand{\zetao}{\zeta_0}

\newcommand{\cstar}{c^\ast}

\newcommand{\phibar}{\overline{\phi}}

\newcommand{\cs}{\mathrm{cs}}
\renewcommand{\sc}{\mathrm{sc}}
\newcommand{\FE}{\Theta}
\newcommand{\fs}{\textrm{max}}

\newcommand{\tNIp}{\widetilde{N}_{1,\,\mathrm{p}}}

\newcommand{\teetapve}{\widetilde{\eta}_{\mathrm{p},\,\ast}}
\newcommand{\teetapfe}{\widetilde{\eta}_\mathrm{p,\,fe}}
\newcommand{\teetap}{\widetilde{\eta}_\mathrm{p}}
\newcommand{\teetapo}{\widetilde{\eta}_\mathrm{p,\,0}}
\providecommand{\eps}{\varepsilon}

\providecommand{\UI}{U_{1}}

\providecommand{\Deve}{\mathrm{De}_{\ast}}
\providecommand{\Devebar}{\overline{\mathrm{De}}_{\ast}}
\providecommand{\Devebarpred}{\overline{\mathrm{De}}_{\ast,\,\mathrm{pred}}}

\providecommand{\Kstar}{K_{\ast}}

%% file: main_v4.1_jor.bbl
\begin{thebibliography}{59}
\providecommand{\natexlab}[1]{#1}
\providecommand{\url}[1]{\texttt{#1}}
\expandafter\ifx\csname urlstyle\endcsname\relax
  \providecommand{\doi}[1]{doi: #1}\else
  \providecommand{\doi}{doi: \begingroup \urlstyle{rm}\Url}\fi

\bibitem[{A. Peterlin}({1966})]{Peterlin1966b}
{A. Peterlin}.
\newblock {Hydrodynamics of macromolecules in a velocity field with
  longitudinal gradient}.
\newblock \emph{{J. Poly. Sci. B, Poly. Lett.}}, {4B}:\penalty0 {287--291},
  {1966}.

\bibitem[Aisling et~al.(2024)Aisling, Saraka, and Alvarez]{Aisling2024-ie}
Ann Aisling, Renee Saraka, and Nicolas~J Alvarez.
\newblock The importance of initial extension rate on elasto-capillary thinning
  of dilute polymer solutions.
\newblock \emph{J. Non-Newton. Fluid Mech.}, 333:\penalty0 105321, November
  2024.
\newblock \doi{10.1016/j.jnnfm.2024.105321}.

\bibitem[Anna and McKinley(2001)]{annamckinley}
S.~L. Anna and G.~H. McKinley.
\newblock Elasto-capillary thinning and breakup of model elastic liquids.
\newblock \emph{J. Rheol.}, 45:\penalty0 115--138, 2001.

\bibitem[Bazilevsky et~al.(1990)Bazilevsky, Entov, and Rozhkov]{bazilevsky1}
A.~V. Bazilevsky, V.~M. Entov, and A.~N. Rozhkov.
\newblock Liquid filament microrheometer and some of its applications.
\newblock In D.~R. Oliver, editor, \emph{Proc. Third Euro. Rheol. Conf.}, pages
  41--43, 1990.

\bibitem[Bhattacharjee et~al.(2011)Bhattacharjee, Mc{D}onnell, Prabhakar, Yeo,
  and Friend]{Bhattacharjee2011}
P.~K. Bhattacharjee, A.~G. Mc{D}onnell, R.~Prabhakar, L.~Y. Yeo, and J.~Friend.
\newblock Extensional flow of low-viscosity fluids in capillary bridges formed
  by pulsed surface acoustic wave jetting.
\newblock \emph{New J. Phys.}, 13:\penalty0 023005, 2011.

\bibitem[Bird et~al.(1987{\natexlab{a}})Bird, Curtiss, Armstrong, and
  Hassager]{dpl1}
R.~B. Bird, C.~F. Curtiss, R.~C. Armstrong, and O.~Hassager.
\newblock \emph{Dynamics of Polymeric Liquids}, volume 1. {F}luid mechanics.
\newblock Wiley-Interscience, New York, 2 edition, 1987{\natexlab{a}}.

\bibitem[Bird et~al.(1987{\natexlab{b}})Bird, Curtiss, Armstrong, and
  Hassager]{dpl2}
R.~B. Bird, C.~F. Curtiss, R.~C. Armstrong, and O.~Hassager.
\newblock \emph{Dynamics of Polymeric Liquids}, volume 2. {K}inetic theory.
\newblock Wiley-Interscience, New York, 2 edition, 1987{\natexlab{b}}.

\bibitem[Buchholz et~al.(2004)Buchholz, Zahn, Kenward, Slater, and
  Barron]{Buchholz2004-mh}
Brett~A Buchholz, Jacob~M Zahn, Martin Kenward, Gary~W Slater, and Annelise~E
  Barron.
\newblock Flow-induced chain scission as a physical route to narrowly
  distributed, high molar mass polymers.
\newblock \emph{Polymer (Guildf.)}, 45\penalty0 (4):\penalty0 1223--1234,
  February 2004.

\bibitem[Calabrese et~al.(2025)Calabrese, Shen, and Haward]{Calabrese2025-re}
V~Calabrese, A~Q Shen, and S~J Haward.
\newblock Effects of polydispersity and concentration on elastocapillary
  thinning of dilute polymer solutions.
\newblock \emph{Phys. Rev. X.}, 15\penalty0 (2):\penalty0 021025, April 2025.

\bibitem[Campo-Deano and Clasen({2010})]{CDeano2010}
L.~Campo-Deano and C.~Clasen.
\newblock {The slow retraction method (SRM) for the determination of
  ultra-short relaxation times in capillary breakup extensional rheometry
  experiments}.
\newblock \emph{{J. Non-Newtonian Fluid Mech.}}, {165}:\penalty0 {1688--1699},
  {2010}.

\bibitem[Clasen et~al.(2006)Clasen, Plog, Kulicke, Owens, Macosko, Scriven,
  Verani, and McKinley]{clasenetal}
C.~Clasen, J.~P. Plog, W.~M. Kulicke, M.~Owens, C.~Macosko, L.E. Scriven,
  M.~Verani, and G.H. McKinley.
\newblock How dilute are dilute solutions in extensional flows?
\newblock \emph{J. Rheol.}, 50:\penalty0 849--881, 2006.

\bibitem[Deblais et~al.(2020)Deblais, Herrada, Eggers, and
  Bonn]{Deblais2020-of}
A~Deblais, M~A Herrada, J~Eggers, and D~Bonn.
\newblock Self-similarity in the breakup of very dilute viscoelastic solutions.
\newblock \emph{J. Fluid Mech.}, 904:\penalty0 R2, December 2020.

\bibitem[Dinic et~al.(2015)Dinic, Zhang, Jimenez, and Sharma]{Dinic2015-fh}
Jelena Dinic, Yiran Zhang, Leidy~Nallely Jimenez, and Vivek Sharma.
\newblock Extensional relaxation times of dilute, aqueous polymer solutions.
\newblock \emph{ACS Macro Lett.}, 4\penalty0 (7):\penalty0 804--808, July 2015.

\bibitem[Doi and Edwards(1986)]{doiedw}
M.~Doi and S.~F. Edwards.
\newblock \emph{{The Theory of Polymer Dynamics}}.
\newblock {Oxford University Press}, 1986.

\bibitem[Ebagninin et~al.(2009)Ebagninin, Benchabane, and
  Bekkour]{Ebagninin2009}
Koblan~Wilfried Ebagninin, Adel Benchabane, and Karim Bekkour.
\newblock Rheological characterization of poly(ethylene oxide) solutions of
  different molecular weights.
\newblock \emph{J. Colloid Interface Sci.}, 336\penalty0 (1):\penalty0
  360--367, 2009.

\bibitem[Eggers(1997)]{eggers1997}
J.~Eggers.
\newblock Nonlinear dynamics and breakup of free-surface flows.
\newblock \emph{Rev. Mod. Phys.}, 69:\penalty0 865--930, 1997.

\bibitem[Eggers et~al.(2020)Eggers, Herrada, and Snoeijer]{Eggers2020-rn}
J~Eggers, M~A Herrada, and J~H Snoeijer.
\newblock Self-similar breakup of polymeric threads as described by the
  oldroyd-{B} model.
\newblock \emph{J. Fluid Mech.}, 887\penalty0 (A19):\penalty0 A19, March 2020.

\bibitem[Entov and Hinch(1997)]{entovhinch}
V.~M. Entov and E.~J. Hinch.
\newblock Effect of a spectrum of relaxation times on the capillary thinning of
  a filament of elastic liquid.
\newblock \emph{J. Non-Newtonian Fluid Mech.}, 72:\penalty0 31--53, 1997.

\bibitem[Ewoldt and McKinley(2017)]{Ewoldt2017-sx}
Randy~H Ewoldt and Gareth~H McKinley.
\newblock Mapping thixo-elasto-visco-plastic behavior.
\newblock \emph{Rheol. Acta}, 56\penalty0 (3):\penalty0 195--210, March 2017.

\bibitem[Gaillard et~al.(2024)Gaillard, Herrada, Deblais, Eggers, and
  Bonn]{Gaillard2024-xu}
A~Gaillard, M~A Herrada, A~Deblais, J~Eggers, and D~Bonn.
\newblock Beware of {CaBER}: Filament thinning rheometry does not always give
  ‘the’ relaxation time of polymer solutions.
\newblock \emph{Phys. Rev. Fluids}, 9\penalty0 (7), July 2024.

\bibitem[Gaillard et~al.(2025)Gaillard, Herrada, Deblais, van Poelgeest,
  Laruelle, Eggers, and Bonn]{Gaillard2025-on}
A~Gaillard, M~A Herrada, A~Deblais, C~van Poelgeest, L~Laruelle, J~Eggers, and
  D~Bonn.
\newblock When does the elastic regime begin in viscoelastic pinch-off?
\newblock \emph{J. Fluid Mech.}, 1005:\penalty0 A10, 2025.
\newblock \doi{10.1017/jfm.2024.1222}.

\bibitem[Gupta et~al.(2000)Gupta, Nguyen, and Sridhar]{gupta}
R.~K. Gupta, D.~A. Nguyen, and T.~Sridhar.
\newblock Extensional viscosity of dilute polystyrene solutions: Effect of
  concentration and molecular weight.
\newblock \emph{Phys. Fluids}, 12:\penalty0 1296--1318, 2000.

\bibitem[Hu et~al.(2025)Hu, Hwang, Ruangkriengsin, and Stone]{Hu2025-rt}
Nan Hu, Jonghyun Hwang, Tachin Ruangkriengsin, and Howard~A Stone.
\newblock Revealing actual viscoelastic relaxation times in capillary breakup.
\newblock \emph{Phys. Rev. Lett.}, 135\penalty0 (4):\penalty0 048201, July
  2025.
\newblock \doi{10.1103/2jz7-4w4k}.

\bibitem[Joseph and Rothstein(2025)]{Joseph2025-ls}
Joe~B Joseph and Jonathan~P Rothstein.
\newblock Recovery dynamics and polymer scission in capillary breakup
  extensional rheometry.
\newblock \emph{J. Nonnewton. Fluid Mech.}, 337\penalty0 (105396):\penalty0
  105396, March 2025.

\bibitem[Kishbaugh and McHugh(1990)]{Kishbaugh1990}
A.~J. Kishbaugh and A.~J. McHugh.
\newblock {A discussion of shear-thickening in bead-spring models}.
\newblock \emph{J Non-Newton. Fluid Mechanics}, 34:\penalty0 181--206, 1990.

\bibitem[Kr{\"o}ger et~al.(2000)Kr{\"o}ger, Alba-P{\'e}rez, Laso, and
  {\"O}ttinger]{Kroger2000}
M~Kr{\"o}ger, A~Alba-P{\'e}rez, M~Laso, and H~C {\"O}ttinger.
\newblock {Variance reduced Brownian simulation of a bead-spring chain under
  steady shear flow considering hydrodynamic interaction effects}.
\newblock \emph{J. Chem. Phys.}, 113:\penalty0 4767--4773, 2000.

\bibitem[Larson(2005)]{larsonreview}
R.~G. Larson.
\newblock The rheology of dilute solutions of flexible polymers: {P}rogress and
  problems.
\newblock \emph{J. Rheol.}, 49:\penalty0 1--70, 2005.

\bibitem[Lee and Muller(1999)]{Lee1999-pj}
Ellen~C Lee and Susan~J Muller.
\newblock Flow light scattering studies of polymer coil conformation in
  solutions under shear: effect of solvent quality.
\newblock \emph{Polymer}, 40\penalty0 (10):\penalty0 2501--2510, 1~May 1999.

\bibitem[Li and Sprittles(2016)]{li_sprittles_2016}
Y.~Li and J.~E. Sprittles.
\newblock Capillary breakup of a liquid bridge: identifying regimes and
  transitions.
\newblock \emph{Journal of Fluid Mechanics}, 797:\penalty0 29–59, 2016.

\bibitem[Liang and Mackley(1994)]{liang}
R.~F. Liang and M.~R. Mackley.
\newblock Rheological characterization of the time and strain dependence for
  polyisobutylene solutions.
\newblock \emph{J. Non-Newtonian Fluid Mech.}, 52:\penalty0 387--405, 1994.

\bibitem[Magda et~al.(1988)Magda, Larson, and Mackay]{magda}
J.~J. Magda, R.~G. Larson, and M.~E. Mackay.
\newblock Deformation-dependent hydrodynamic interaction in flows of dilute
  polymer solutions.
\newblock \emph{J. Chem. Phys.}, 89:\penalty0 2504--2513, 1988.

\bibitem[McDonnell et~al.(2015)McDonnell, Gopesh, Lo, O'Bryan, Yeo, Friend, and
  Prabhakar]{mcdonnell2015}
A.~G. McDonnell, T.~C. Gopesh, J.~Lo, M.~O'Bryan, L.~Y. Yeo, J.~R. Friend, and
  R.~Prabhakar.
\newblock Motility induced changes in viscosity of suspensions of swimming
  microbes in extensional flows.
\newblock \emph{Soft Matter}, 11:\penalty0 4658--4668, 2015.

\bibitem[McKinley(2005)]{McKinley2005-fl}
G~H McKinley.
\newblock Visco-elasto-capillary thinning and break-up of complex fluids.
\newblock Technical report, Cambridge, MA, 2005.

\bibitem[McKinley and Sridhar(2002)]{mckinleysridhar}
G.~H. McKinley and T.~Sridhar.
\newblock Filament-stretching rheometry of complex fluids.
\newblock \emph{Ann. Rev. Fluid Mech.}, 34:\penalty0 375--415, 2002.

\bibitem[McKinley and Tripathi(2000)]{McKinley2000}
G.~H. McKinley and A.~Tripathi.
\newblock {How to extract the Newtonian viscosity from capillary breakup
  measurements in a filament rheometer}.
\newblock \emph{J. Rheol.}, 44:\penalty0 653--670, 2000.

\bibitem[Odell et~al.(1990)Odell, Muller, Narh, and Keller]{Odell1990-so}
Jeffrey~A Odell, Alejandro~J Muller, Kwabena~A Narh, and Andrew Keller.
\newblock Degradation of polymer solutions in extensional flows.
\newblock \emph{Macromolecules}, 23\penalty0 (12):\penalty0 3092--3103, June
  1990.

\bibitem[{\"{O}}ttinger(1987)]{Ottinger87}
H.~C. {\"{O}}ttinger.
\newblock Generalized {Z}imm model for dilute polymer solutions under theta
  conditions.
\newblock \emph{J. Chem. Phys.}, 86:\penalty0 3731--3749, 1987.

\bibitem[{\"{O}}ttinger(1989)]{Ottinger89}
H.~C. {\"{O}}ttinger.
\newblock Gaussian approximation for {R}ouse chains with hydrodynamic
  interaction.
\newblock \emph{J. Chem. Phys.}, 90:\penalty0 463--473, 1989.

\bibitem[{\"{O}}ttinger and Zylka(1992)]{Ottinger92}
H.~C. {\"{O}}ttinger and W.~Zylka.
\newblock On the relaxation spectra for models of dilute polymer solutions.
\newblock \emph{J. Rheol.}, 36:\penalty0 885--910, 1992.

\bibitem[Papageorgiou(1995)]{papageorgiou1995}
D.~Papageorgiou.
\newblock {Analytical description of the breakup of liquid jets}.
\newblock \emph{J. Fluid Mech.}, 301:\penalty0 109--132, 1995.

\bibitem[Pipkin(1986)]{Pipkin1986-wg}
Allen~C Pipkin.
\newblock \emph{Lectures on Viscoelasticity Theory}.
\newblock Applied Mathematical Sciences. Springer, New York, NY, 2 edition,
  June 1986.

\bibitem[Prabhakar(2026{\natexlab{a}})]{c2d2jl}
R.~Prabhakar.
\newblock {C2D2.jl}: Julia code for the conformation- and
  concentration-dependent drag (c2d2) model in steady and unsteady homogeneous
  flows, July 2026{\natexlab{a}}.
\newblock URL \url{https://doi.org/10.5281/zenodo.21090147}.
\newblock Software (MIT License), version v1.0.0, commit c4f6be4. Source:
  https://github.com/prabhakarranganathan/C2D2.jl.

\bibitem[Prabhakar(2026{\natexlab{b}})]{c2d2report}
R.~Prabhakar.
\newblock \emph{Technical Note on the Conformation- and Concentration-Dependent
  Drag (C2D2) model for unentangled solutions of flexible polymers and
  numerical algorithms for rheological properties in steady and unsteady
  flows.}, July 2026{\natexlab{b}}.
\newblock URL \url{https://doi.org/10.5281/zenodo.18012677}.
\newblock Concept DOI; resolves to the latest version.

\bibitem[Prabhakar and Prakash(2006)]{prabhakar_gaussian}
R.~Prabhakar and J.~R. Prakash.
\newblock Gaussian approximation for finitely extensible bead-spring chains
  with hydrodynamic interaction.
\newblock \emph{J. Rheol.}, 50:\penalty0 561--593, 2006.

\bibitem[Prabhakar et~al.(2004)Prabhakar, Prakash, and Sridhar]{prabhakarSFG}
R.~Prabhakar, J.~R. Prakash, and T.~Sridhar.
\newblock A successive fine-graining scheme for predicting the rheological
  properties of dilute polymer solutions.
\newblock \emph{J. Rheol.}, 48:\penalty0 1251--1278, 2004.

\bibitem[Prabhakar et~al.(2006)Prabhakar, Prakash, and Sridhar]{prabhakar}
R.~Prabhakar, J.~R. Prakash, and T.~Sridhar.
\newblock Effect of configuration-dependent intramolecular hydrodynamic
  interaction on elastocapillary thinning and breakup of filaments of dilute
  polymer solutions.
\newblock \emph{J. Rheol.}, 50:\penalty0 925--947, 2006.

\bibitem[Prabhakar et~al.(2016)Prabhakar, Gadkari, Gopesh, and
  Shaw]{Prabhakar2016}
R.~Prabhakar, S.~Gadkari, T.~Gopesh, and M.~J. Shaw.
\newblock Influence of stretching induced self-concentration and self-dilution
  on coil-stretch hysteresis and capillary thinning of unentangled polymer
  solutions.
\newblock \emph{J. Rheol.}, 60:\penalty0 345--366, 2016.

\bibitem[Prabhakar et~al.(2017)Prabhakar, Sasmal, Nguyen, Sridhar, and
  Prakash]{Prabhakar2017-ri}
R.~Prabhakar, C.~Sasmal, D.~A. Nguyen, T.~Sridhar, and J.~R. Prakash.
\newblock Effect of stretching-induced changes in hydrodynamic screening on
  coil-stretch hysteresis of unentangled polymer solutions.
\newblock \emph{Phys. Rev. Fluids}, 2\penalty0 (1):\penalty0 011301, January
  2017.

\bibitem[Prabhakar and Connell(2026)]{SI}
Ranganathan Prabhakar and Joseph Connell.
\newblock Supplementary information, 2026.
\newblock Supplementary Information accompanying the manuscript.

\bibitem[Robertson and Calabrese(2022)]{Robertson_Calabrese}
Benjamin Robertson and Michelle Calabrese.
\newblock Evaporation-controlled dripping-onto-substrate (dos) extensional
  rheology of viscoelastic polymer solutions.
\newblock \emph{Scientific Reports}, 12, 03 2022.
\newblock \doi{10.1038/s41598-022-08448-x}.

\bibitem[Rubinstein and Colby(2003)]{rubinsteincolby}
M.~Rubinstein and R.~H. Colby.
\newblock \emph{Polymer physics}.
\newblock Oxford University Press, London, UK, 2003.

\bibitem[Sasmal et~al.(2017)Sasmal, Hsiao, Schroeder, and
  Ravi~Prakash]{Sasmal2016}
Chandi Sasmal, Kai-Wen Hsiao, Charles~M Schroeder, and J~Ravi~Prakash.
\newblock Parameter-free prediction of {DNA} dynamics in planar extensional
  flow of semidilute solutions.
\newblock \emph{J. Rheol.}, 61\penalty0 (1):\penalty0 169--186, 1~January 2017.

\bibitem[Solomon and Muller(1996)]{Solomon1996b}
M.~J. Solomon and S.~J. Muller.
\newblock Study of mixed solvent quality in a polystyrene--dioctyl
  phthalate--polystyrene system.
\newblock \emph{J. Polym. Sci. B Polym. Phys.}, 34\penalty0 (1):\penalty0
  181--192, 1996.

\bibitem[Tirtaatmadja and Sridhar(1993)]{Tirtaatmadja1993}
V.~Tirtaatmadja and T.~Sridhar.
\newblock A filament stretching device for measurement of extensional
  viscosity.
\newblock \emph{J. Rheol.}, 37:\penalty0 1081--1102, 1993.

\bibitem[Tirtaatmadja et~al.(2006)Tirtaatmadja, McKinley, and
  Cooper-White]{Tirtaatmadja2006}
V.~Tirtaatmadja, G.~H. McKinley, and J.~J. Cooper-White.
\newblock Drop formation and breakup of low viscosity elastic fluids: Effects
  of molecular weight and concentration.
\newblock \emph{Phys. Fluids}, 18:\penalty0 043101, 2006.

\bibitem[Tuladhar and Mackley(2008)]{Tuladhar2008-nq}
T~R Tuladhar and M~R Mackley.
\newblock Filament stretching rheometry and break-up behaviour of low viscosity
  polymer solutions and inkjet fluids.
\newblock \emph{J. Non-Newtonian Fluid Mech.}, 148\penalty0 (1-3):\penalty0
  97--108, January 2008.

\bibitem[Vadillo et~al.(2010)Vadillo, Tuladhar, Mulji, Jung, Hoath, and
  Mackley]{EvalInkJet}
D.~Vadillo, T.~Tuladhar, A.~C. Mulji, S.~Jung, S.~Hoath, and M.~R. Mackley.
\newblock Evaluation of the inkjet fluid’s performance using the “cambridge
  trimaster” filament stretch and break-up device.
\newblock \emph{Journal of Rheology}, 54, 2010.

\bibitem[Wang et~al.(2025)Wang, Prabhakar, Ravi~Prakash, and
  Larson]{Wang2025-sf}
Hao-Yu Wang, Ranganathan Prabhakar, J~Ravi~Prakash, and Ronald~G Larson.
\newblock Using brownian dynamics simulations to demystify capillary breakup
  extensional rheometry ({CaBER}).
\newblock \emph{J. Rheol. (N. Y. N. Y.)}, 69\penalty0 (5):\penalty0 641--656,
  1~September 2025.

\bibitem[Zhang and Calabrese(2022)]{Zhang_Calabrese}
Diana Zhang and Michelle Calabrese.
\newblock Temperature-controlled dripping-onto-substrate (dos) extensional
  rheometry of polymer micelle solutions.
\newblock \emph{Soft Matter}, 18, 05 2022.
\newblock \doi{10.1039/D2SM00377E}.

\end{thebibliography}
